\documentclass[11pt,a4paper]{article}
\pdfoutput=1
\usepackage{jheppub}
\usepackage{tikz}
\usetikzlibrary{intersections, calc, arrows.meta}
\usepackage{subcaption}
\usepackage{bm}

\DeclareMathOperator{\Tr}{Tr}

\newcommand{\ri}{\mathrm{i}}
\renewcommand{\th}{\theta}
\newcommand{\Th}{\Theta}
\newcommand{\cob}{\delta}

\newcommand{\hf}{\frac{1}{2}}
\newcommand{\qu}{\frac{1}{4}}
\newcommand{\til}[1]{\widetilde{#1}}
\newcommand{\si}{\sigma}

\newcommand{\del}{\partial}

\newcommand{\lap}{\Delta}
\newcommand{\bra}{\langle}
\newcommand{\ket}{\rangle}
\newcommand{\la}{\lambda}

\newcommand{\ka}{\kappa}

\newcommand{\bt}{\beta}

\newcommand{\Ga}{\Gamma}
\newcommand{\al}{\alpha}

\newcommand{\rt}[1]{\sqrt{#1}}
\newcommand{\cO}{\mathcal{O}}

\newcommand{\cF}{\mathcal{F}}

\newcommand{\cG}{\mathcal{G}}
\newcommand{\cL}{\mathcal{L}}
\newcommand{\cB}{\mathcal{B}}

\newcommand{\cN}{\mathcal{N}}

\newcommand{\mZ}{\mathbb{Z}}
\newcommand{\mat}[1]{\begin{pmatrix}#1\end{pmatrix}}

\begin{document}
\preprint{RUP-22-18}

\title{Large $N$ expansion of an integrated correlator in $\cN=4$ SYM}

\author[a]{Yasuyuki Hatsuda}
\author[b]{and Kazumi Okuyama}

\affiliation[a]{Department of Physics, Rikkyo University, Toshima, Tokyo 171-8501, Japan}
\affiliation[b]{Department of Physics, 
Shinshu University, 3-1-1 Asahi, Matsumoto 390-8621, Japan}

\emailAdd{yhatsuda@rikkyo.ac.jp, kazumi@azusa.shinshu-u.ac.jp}

\abstract{
Recently Dorigoni, Green and Wen conjectured a remarkable exact formula for an integrated correlator of four superconformal primary operators in $\mathcal{N}=4$ supersymmetric Yang-Mills theory. In this work, we investigate its large $N$ limit in detail. We show that the formula of Dorigoni, Green and Wen can be recast into the sum over the
contributions of $(p,q)$-strings. Due to the $SL(2,\mZ)$ duality, all the contributions are governed by a single function, typically appearing as the fundamental string contribution. The large order behavior for the perturbative genus expansion of this function allows us to reveal the large $N$ non-perturbative corrections,
which we interpret as the D3-brane instantons in the holographically dual
type IIB string theory.
The same result is obtained more systematically by using a Laplace-difference equation for the integrated correlator.} 

\maketitle

\section{Introduction}
The holographic duality 
between 4d $\cN=4$ $SU(N)$ supersymmetric Yang-Mills theory (SYM)
and the type IIB string theory on $AdS_5\times S^5$
is a prototypical example of the AdS/CFT correspondence \cite{Maldacena:1997re}. 
In particular, Montonen-Olive duality of $\cN=4$
SYM \cite{Montonen:1977sn,Witten:1978mh} 
corresponds to the $SL(2,\mZ)$ duality of type IIB string theory
\cite{Hull:1994ys}.
However, implications of the 
$SL(2,\mZ)$ duality in the context of AdS/CFT correspondence have
not been fully explored.
This is partly because, in the large $N$ 't Hooft limit
\begin{equation}
\begin{aligned}
 N\to\infty,\qquad g_{\text{YM}}\to0,\qquad\la=g_{\text{YM}}^2N:\text{fixed},
\end{aligned} 
\label{eq:'tHooft}
\end{equation}
we are focusing on the small $g_{\text{YM}}$ regime and 
the $SL(2,\mZ)$ duality is obscured in this limit.
However, one can study the holography and the $SL(2,\mZ)$ duality
at the same time in a regime where
\begin{equation}
\begin{aligned}
 N\to\infty,\qquad g_{\text{YM}}:\text{fixed},
\end{aligned} 
\label{eq:vsc}
\end{equation}
which is called ``very strong coupling limit'' in \cite{Azeyanagi:2013fla}.

In recent remarkable papers \cite{Dorigoni:2021bvj,Dorigoni:2021guq}, 
Dorigoni, Green and Wen conjectured an exact form
of an integrated four-point correlator $\cG_N$ of 
superconformal primary operators in the ${\bf 20'}$ 
of the $SU(4)_R$ symmetry of $\cN=4$ SYM. Such integrated correlators were originally introduced in \cite{Binder:2019jwn} to analyze the strong coupling limit of usual four-point functions. See \cite{Chester:2019pvm,Chester:2020dja,Chester:2019jas,Chester:2020vyz,Collier:2022emf,Dorigoni:2022zcr,Dorigoni:2022iem} for the related works on the integrated correlators
in $\cN=4$ SYM.
The result in \cite{Dorigoni:2021bvj,Dorigoni:2021guq} is written in terms of the 
sum over the two-dimensional lattice $m+n\tau~((m,n)\in\mZ^2)$
where $\tau$ is the complexified coupling of $\cN=4$ SYM.
It is manifestly invariant under the $SL(2,\mZ)$ duality.
They computed the worldsheet instanton correction of the form
$e^{-2\rt{\la}}$ in the large $N$ 't Hooft limit.

In \cite{Collier:2022emf}, Collier and Perlmutter further studied
this integrated correlator $\cG_N$
using the spectral decomposition 
in terms of the eigenfunctions of the Laplacian on the
upper-half $\tau$-plane with hyperbolic metric.
They found that $\cG_N$ receives
an additional non-perturbative correction of the form $e^{-2\rt{\la}/g_s}$,
where $g_s=g_{\text{YM}}^2/4\pi$ is the string coupling of 
the bulk type IIB theory;
they identified this non-perturbative correction
as the ``D-string instanton''.

In this paper, we will study the large $N$
expansion of the integrated correlator $\cG_N$ in detail.
We first show that the lattice sum in \cite{Dorigoni:2021bvj,Dorigoni:2021guq} can be recast into the form of the sum over the
contribution of $(p,q)$-strings. All of these contributions are symmetric.
This is consistent with the expectation 
from the $SL(2,\mZ)$ duality of bulk type IIB string theory.
Then we focus on the contribution of the $(1,0)$-string,
which can be written as the usual genus expansion 
$\sum_g N^{2-2g}G_g(\la)$ with the fixed 't Hooft coupling
$\la$.
However, as noticed by Collier and Perlmutter,
the $(1,0)$-string contribution has another representation, 
which can also be written as the genus-like expansion 
$\sum_g N^{1-2g}\til{G}_g(\til{\la})$ in the dual 't Hooft limit,
\begin{equation}
\begin{aligned}
 N\to\infty,\qquad g_{\text{YM}}\to\infty,\qquad\til{\la}=\frac{16\pi^2}{g_{\text{YM}}^2} N:\text{fixed}.
\end{aligned} 
\label{eq:dual'tHooft}
\end{equation}
Both of these genus expansions are determined by
a single spectral function $f_g(s)$.
From the Laplace-difference equation for 
$\cG_N$ \cite{Dorigoni:2021guq}, we find the recursion relation for $f_g(s)$
which enables us to compute the genus expansion up to very high genera. 
Using these data, we find that the genus expansion of $\cG_N$ is Borel non-summable.
From the large $g$ behavior of $G_g(\la)$ and $\til{G}_g(\til{\la})$,
we find that there are non-perturbative 
correction of the form $e^{-NA(\pi\la^{-1/2})}$ and
$e^{-NA(\pi\til{\la}^{-1/2})}$, 
where the instanton action $A(x)$
has the same form with the action of D3-brane which is holographically 
dual to the so-called ``giant Wilson loop'' in the large symmetric representation
\cite{Drukker:2005kx,Gomis:2006sb,Gomis:2006im}.

\begin{figure}[tb]
\centering
 \includegraphics[width=0.5\linewidth]{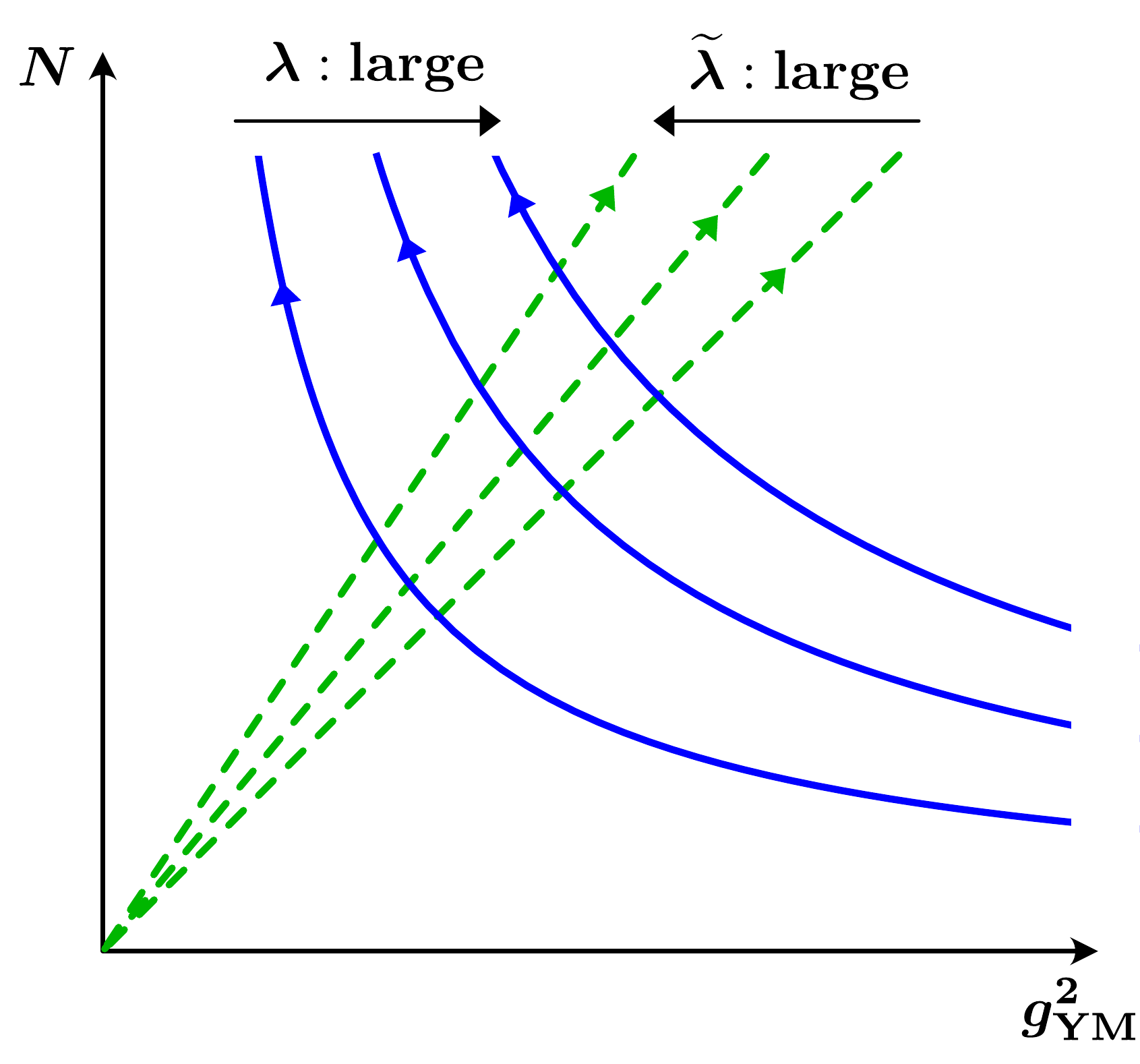}
\caption{The blue solid lines show slices in the 't Hooft limit, and the green dashed lines show slices in the dual 't Hooft limit. For large but finite $N$, the strong 't Hooft coupling limit meets the strong dual 't Hooft coupling limit in the very strong coupling regime. }
\label{fig:N-gs}
\end{figure}

In the usual 't Hooft limit \eqref{eq:'tHooft}, the exponential factor $e^{-NA(\pi\la^{-1/2})}$ is purely non-perturbative in $1/N$.
However, if one considers the dual 't Hooft limit \eqref{eq:dual'tHooft}, 
this correction can be expressed as $e^{-2\til{\la}^{1/2}}$ in terms of $\til{\la}$. Such a correction is non-perturbative in $1/\til{\la}^{1/2}$, but not in $1/N$, and it is visible at each genus in the dual 't Hooft limit.
In fact, we find that the non-perturbative correction $e^{-NA(\pi\la^{-1/2})}$ in this limit reproduces the D-string instanton in \cite{Collier:2022emf}. 
We stress that our non-perturbative correction is originally derived in the 't Hooft limit. However, for very large but finite ($N, \la, \til{\la})$, this can be regarded effectively as the result in an overlapped regime of the 't Hooft limit and the dual 't Hooft limit via the very strong coupling limit. We illustrate it in Figure~\ref{fig:N-gs}. As a consequence, we can extrapolate our result to the dual 't Hooft regime.
Also, the non-perturbative correction $e^{-NA(\pi\til{\la}^{-1/2})}$ reduces to the worldsheet instanton by taking the 't Hooft limit.

This paper is organized as follows.
In section \ref{sec:pq}, we start with a brief review on the integrated correlator $\cG_N$ in $\cN=4$ SYM. We rewrite the integral representation of $\cG_N$ in \cite{Dorigoni:2021bvj,Dorigoni:2021guq} as a sum over $(p,q)$-string contributions. The resulting sum is governed by only a single function, the fundamental string contribution. All the $(p,q)$-string contributions appear democratically with an equal weight. This is a consequence of the $SL(2,\mathbb{Z})$ duality.

In section \ref{sec:spectral}, we take a deep look at the spectral function $f_g(s)$. This function determines both of the genus functions $G_g(\la)$ and $\til{G}_g(\til{\la})$ simultaneously. We find a recursion relation for $f_g(s)$. It enables us to compute the large $N$ expansions up to very high genera.

In section \ref{sec:WS-D1-inst}, we consider the genus expansions of $\cG_N$, and see that the functions $G_g(\la)$ and $\til{G}_g(\til{\la})$ receive non-perturbative corrections in the large (dual) 't Hooft coupling limit. The former is referred to as the worldsheet instanton, and the latter as the D-string instanton in \cite{Collier:2022emf}.

In section \ref{sec:D3-inst}, we explore non-perturbative effects at large $N$. We show that the genus expansions 
$\sum_g N^{2-2g}G_g(\la)$ and $\sum_g N^{1-2g}\til{G}_g(\til{\la})$ receive non-perturbative corrections, which we interpret as the magnetic and electric
D3-instanton corrections, respectively.
From the large order behavior, we constrain the form of the non-perturbative corrections. Interestingly, the same corrections are fixed by the Laplace-difference equation that the integrated correlator obeys. We observe that the dual 't Hooft limit of the non-perturbative correction to $\sum_g N^{2-2g}G_g(\la)$ agrees with the D-string instanton correction, computed in section \ref{sec:WS-D1-inst}. Also the 't Hooft limit of the non-perturbative correction to $\sum_g N^{1-2g}\til{G}_g(\til{\la})$ coincides with the worldsheet instanton correction.

Finally, we conclude in section \ref{sec:conclusion}
with some discussion on the future problems.

In appendix \ref{app:exact}, we prove the relation
between $B_N(t)$ and the matrix model correlator
$K_N(x)$ conjectured in \cite{Dorigoni:2022iem}.
In appendix \ref{app:dis-conn}, we show that
the disconnected and the connected part of $\cG_N$
correspond to $G_0(\la)$ and $\til{G}_0(\til{\la})$ in the large $N$ limit.
In appendix \ref{app:richardson} we summarize the method of 
Richardson extrapolation used in the main text. 

\section{Sum over $(p,q)$-strings}\label{sec:pq}

We start by reviewing the integrated correlator
$\cG_N$ in \cite{Dorigoni:2021guq}, and then rewrite it as a sum over $(p,q)$-string contributions.
$\cG_N$ is written as the derivative of 
the free energy of mass deformed $\cN=2^*$ theory on $S^4$ \cite{Binder:2019jwn} 
\begin{equation}
\begin{aligned}
 \cG_N=\qu\lap_\tau\del_m^2\log Z\Big|_{m=0},
\end{aligned} 
\label{eq:def-GN}
\end{equation} 
where $\lap_\tau$ is the hyperbolic Laplacian
on the upper-half $\tau$-plane
\begin{equation}
\begin{aligned}
 \lap_\tau=\tau_2^2\left(\frac{\del^2}{\del\tau_1^2}
+\frac{\del^2}{\del\tau_2^2}\right),
\end{aligned} 
\label{eq:lap-tau}
\end{equation}
and $\tau$ is the complexified coupling of the 
$SU(N)$ $\cN=4$ SYM
\begin{equation}
\begin{aligned}
 \tau=\tau_1+\ri\tau_2=\frac{\th}{2\pi}+\ri\frac{4\pi}{g^2_{\text{YM}}}.
\end{aligned} 
\end{equation}
Via the holographic duality, $\tau_2$ is related to the string coupling $g_s$ 
of bulk type IIB string theory as
\begin{equation}
\begin{aligned}
 \tau_2=g_s^{-1}.
\end{aligned} 
\end{equation}
The partition function $Z$ of $\cN=2^*$ theory on $S^4$
is exactly computed by the supersymmetric localization \cite{Pestun:2007rz}
and the Yang-Mills instanton contributions,
which correspond to D-instantons in the bulk,
are captured by the Nekrasov partition function
\cite{Nekrasov:2002qd,Nekrasov:2003rj}.

After the supersymmetric localization, the partition function $Z$ is reduced to a finite dimensional matrix integral for moduli parameters.
However, it is not easy to extract the full large $N$ expansion from such a matrix integral because its integrand is complicated.
Very surprisingly, in \cite{Dorigoni:2021guq}, a simple exact form of the integrated correlator
$\cG_N$ is put forward
\begin{equation}
\begin{aligned}
 \cG_N=\hf\int_0^\infty dt\, B_N(t)\Th(t,\tau),
\end{aligned} 
\label{eq:cGN}
\end{equation}
where $\Th(t,\tau)$ is given by the lattice sum
\begin{equation}
\begin{aligned}
\Th(t,\tau)&=\sum_{(m,n)\in\mathbb{Z}^2}e^{-tM_{m,n}},\\
\end{aligned} 
\end{equation}
with
\begin{equation}
\begin{aligned}
 M_{m,n}&=\frac{\pi}{\tau_2}|m+n\tau|^2,
\end{aligned} 
\label{eq:Mmn}
\end{equation}
and $B_N(t)$ in \eqref{eq:cGN} is given by
\begin{equation}
\begin{aligned}
 B_N(t)&=\frac{Q_N(t)}{(1+t)^{2N+1}},\\
Q_N(t)&=-\hf N(N-1)(1-t)^{N-1}(1+t)^{N+1}\\
&\times\left[(3-6t+3t^2+8Nt)P_N^{(1,-2)}\left(\frac{1+t^2}{1-t^2}\right)
+\frac{3t^2-3-8Nt}{1+t}P_N^{(1,-1)}\left(\frac{1+t^2}{1-t^2}\right)\right],
\end{aligned} 
\label{eq:BN}
\end{equation}
where $P_n^{(\al,\bt)}(z)$ denotes the Jacobi polynomial.
One can show that $\Th(t,\tau)$
is manifestly $SL(2,\mZ)$ invariant
\begin{equation}
\begin{aligned}
 \Th\left(t,\frac{a\tau+b}{c\tau+d}\right)=\Th(t,\tau),\qquad
\mat{a&b\\c&d} \in SL(2,\mZ),
\end{aligned} 
\end{equation}
and it satisfies
\begin{equation}
\begin{aligned}
 t^{-1}\Th(t^{-1},\tau)=\Th(t,\tau).
\end{aligned} 
\end{equation}
It turns out that this formula is very useful both in the exact finite $N$ computation and in the large $N$ analysis.

Fourier expansion of $\cG_N$ is obtained by performing the Poisson resummation
of $\Th(t,\tau)$.
Using the relation
\begin{equation}
\begin{aligned}
 \sum_{m\in\mathbb{Z}}\cob(x-m)=\sum_{\til{m}\in\mathbb{Z}}e^{-2\pi\ri \til{m} x},
\end{aligned} 
\end{equation}
$\Th(t,\tau)$ is written as
\begin{equation}
\begin{aligned}
  \Th(t,\tau)
&=\sum_{(m,n)\in\mathbb{Z}^2}e^{-\frac{\pi t}{\tau_2}\bigl[(m+n\tau_1)^2+n^2\tau_2^2
\bigr]}\\
&=\int_{-\infty}^\infty dx\sum_{(\til{m},n)\in\mathbb{Z}^2}e^{-2\pi\ri \til{m} x}
 e^{-\frac{\pi t}{\tau_2}\bigl[(x+n\tau_1)^2+n^2\tau_2^2\bigr]}\\
&=\rt{\frac{\tau_2}{t}}
\sum_{(\til{m},n)\in\mathbb{Z}^2}e^{2\pi\ri \til{m} n\tau_1-
\pi\tau_2\bigl(\frac{\til{m}^2}{t}+n^2t\bigr)}\\
&=\frac{1}{\rt{g_st}}\sum_{(\til{m},n)\in\mathbb{Z}^2}e^{\ri\til{m} n\th-\frac{\pi}{g_s}\bigl(\frac{\til{m}^2}{t}+n^2t\bigr)}.
\end{aligned} 
\label{eq:poisson}
\end{equation}
Thus the $k$-instanton contribution is given by
\begin{equation}
\begin{aligned}
 \cG_N^{(k\text{-inst})}=\hf\sum_{\til{m}n=k}\int_0^\infty \frac{dt}{\rt{g_st}}B_N(t)
e^{-\frac{\pi}{g_s}\bigl(\frac{\til{m}^2}{t}+n^2t\bigr)}.
\end{aligned} 
\label{eq:k-inst}
\end{equation}
In particular, the 0-instanton part ($\til{m}=0$ or $n=0$) is exactly given by
\begin{equation}
\begin{aligned}
 \cG_N^{(\text{0-inst})}&=
\hf\int_0^\infty \frac{dt}{\rt{g_st}}B_N(t)
\Bigl[\vartheta_3(e^{-\frac{2\pi}{g_st}})+\vartheta_3(e^{-\frac{2\pi t}{g_s}})\Bigr],
\end{aligned}
\label{eq:0inst-int} 
\end{equation}
where $\vartheta_3(q)$ denotes the Jacobi theta function
\begin{equation}
\begin{aligned}
 \vartheta_3(q)=\sum_{n\in\mZ}q^{\hf n^2}.
\end{aligned} 
\end{equation}
Using the properties of the Jacobi theta function 
\begin{equation}
\begin{aligned}
 \frac{1}{\rt{g_st}}\vartheta_3(e^{-\frac{2\pi}{g_st}})
&=\vartheta_3(e^{-2\pi g_st}),\\
\frac{1}{\rt{g_st}}\vartheta_3(e^{-\frac{2\pi t}{g_s}})
&=\frac{1}{t}\vartheta_3(e^{-\frac{2\pi g_s}{t}}),
\end{aligned} 
\end{equation}
and the symmetry of $B_N(t)$ \cite{Dorigoni:2021guq},
\begin{equation}
\begin{aligned}
 t^{-1}B_N(t^{-1})=B_N(t),
\end{aligned} 
\end{equation}
one can show that the two terms of \eqref{eq:0inst-int}
are equal. Finally, we find
\begin{equation}
\begin{aligned}
 \cG_N^{(\text{0-inst})}&=\int_0^\infty dt\, B_N(t)\vartheta_3(e^{-2\pi g_st})\\
&=\int_0^\infty dt\, B_N(t)\sum_{n\in\mathbb{Z}}e^{-n^2tM_{1,0}}.
\end{aligned} 
\label{eq:0-inst}
\end{equation} 
We can naturally interpret this $\cG_N^{(\text{0-inst})}$
as the contribution of $(1,0)$-string (or F-string).

Noticing that $M_{m,n}$ in \eqref{eq:Mmn}
has the property 
\begin{equation}
\begin{aligned}
 M_{np,nq}=n^2M_{p,q},
\end{aligned} 
\end{equation}
we can rewrite \eqref{eq:cGN} as\footnote{This rewriting is equivalent to
the Poincar\'{e} sum discussed in \cite{Collier:2022emf} (see e.g. \cite{bruinier20081}).
We would like to thank Eric Perlmutter for pointing this out.} 
\begin{equation}
\begin{aligned}
 \cG_N&=\hf\int_0^\infty dt\, B_N(t)\sum_{(m,n)\in\mathbb{Z}^2}e^{-tM_{m,n}}\\
&=\hf\int_0^\infty dt\, B_N(t)\sum_{\substack{p,q\\\text{gcd}(p,q)=1}}\sum_{n\in\mathbb{Z}}e^{-tM_{np,nq}}\\
&=\hf\int_0^\infty dt\, B_N(t)\sum_{\substack{p,q\\\text{gcd}(p,q)=1}}
\sum_{n\in\mathbb{Z}}e^{-n^2t M_{p,q}}\\
&=\hf \sum_{\substack{p,q\\\text{gcd}(p,q)=1}} \cG_N^{\text{(0-inst)}}(M_{p,q}),
\end{aligned} 
\label{eq:pq-sum}
\end{equation}
where we defined
\begin{equation}
\begin{aligned}
 \cG_N^{\text{(0-inst)}}(M):=\int_0^\infty dt\, B_N(t)\sum_{n\in\mathbb{Z}}
e^{-n^2tM}=\int_0^\infty dt\, B_N(t)\vartheta_3(e^{-2Mt}),
\end{aligned} 
\label{eq:0-inst-M}
\end{equation}
which, of course, reduces to the 0-instanton part \eqref{eq:0-inst} when $M=M_{1,0}=\pi g_s$.
Now, we interpret \eqref{eq:pq-sum} as the sum over the contribution of $(p,q)$-strings
with relatively prime $(p,q)$.
In fact, $M_{p,q}$ has a physical meaning as the square of the $(p,q)$-string tension
\cite{Schwarz:1995dk}.
As expected from the $SL(2,\mZ)$ duality invariance of $\cG_N$, all contributions of
the $(p,q)$-strings have the same functional form 
$\cG_N^{\text{(0-inst)}}(M)$ as a function 
of $M=M_{p,q}$.
In other words, all $(p,q)$-strings appear democratically in $\cG_N$
with an equal weight.

In the following sections, we will focus on the zero-instanton term $\cG_N^{\text{(0-inst)}}(M)$. 
This is the only building block of the full function $\cG_N$. 
It turns out that the non-perturbative corrections
in the large $N$ expansion of $\cG_N^{\text{(0-inst)}}(M)$
do not correspond to the D-instanton.
The D-instanton effect arises only after we sum over all $(p,q)$
in \eqref{eq:pq-sum} and perform the Poisson resummation \eqref{eq:poisson}.
In other words, the D-instanton cannot be seen in 
$\cG_N^{\text{(0-inst)}}(M_{p,q})$ with fixed $(p,q)$. 

In the large $N$ limit, only a single term
in the sum \eqref{eq:pq-sum} is dominant.
Which $(p,q)$ becomes dominant is determined by the condition
\begin{equation}
\begin{aligned}
 \min_{(p,q)}M_{p,q}.
\end{aligned} 
\label{eq:pq-min}
\end{equation}
This condition divides the upper-half $\tau$-plane into
several domains.
For instance, the $(1,0)$-string (or F-string)
is dominant in the fundamental domain of $SL(2,\mZ)$.
In general, the condition \eqref{eq:pq-min}
leads to a tessellation of the upper-half $\tau$-plane,
which is similar to the phase diagram
of 3d gravity (see e.g. Figure 3 in \cite{Maloney:2007ud}).

\section{Spectral representation}\label{sec:spectral}
The $N$-dependence of the integrated correlator \eqref{eq:cGN} is encoded in the function $B_N(t)$ in the integrand. 
As discussed in \cite{Dorigoni:2021guq,Collier:2022emf}, it is more convenient
to consider the coefficient $c(s,N)$
appearing in the small $t$ expansion of $B_N(t)$
\begin{equation}
\begin{aligned}
 B_N(t)&=\sum_{s=2}^\infty (-1)^s c(s,N)t^{s-1}.
\end{aligned} 
\label{eq:B-cs}
\end{equation}
Although $c(s,N)$ is only defined for positive integer $s$
in this expression, 
one can naturally define the analytic continuation of 
$c(s,N)$ to the complex $s$-plane, as emphasized 
in \cite{Dorigoni:2021guq,Collier:2022emf}.

In this section, we first consider the large $N$ limit of $c(s,N)$. 
We will see that the usual resummation technique allows us to re-construct the finite $N$ result of $c(s,N)$ from the large $N$ expansion.
In the next section, we will perform the similar analysis for 
$\cG_N^{\text{(0-inst)}}(M_{1,0})$, and it turns out that the large $N$ expansion of $\cG_N^{\text{(0-inst)}}(M_{1,0})$ receives non-perturbative corrections.

\subsection{Genus expansion}
Let us consider $c(s,N)$ defined in \eqref{eq:B-cs}.
For small $N$, we can compute the coefficient $c(s,N)$ explicitly, as in \eqref{eq:c-finite-N-2} for $N=2,3,4$.
It is interesting to notice that one can write down an exact closed form 
of $c(s,N)$ in \eqref{eq:c-finite-N} for finite $N$.

\begin{figure}[tb]
\centering
\includegraphics
[width=0.6\linewidth]{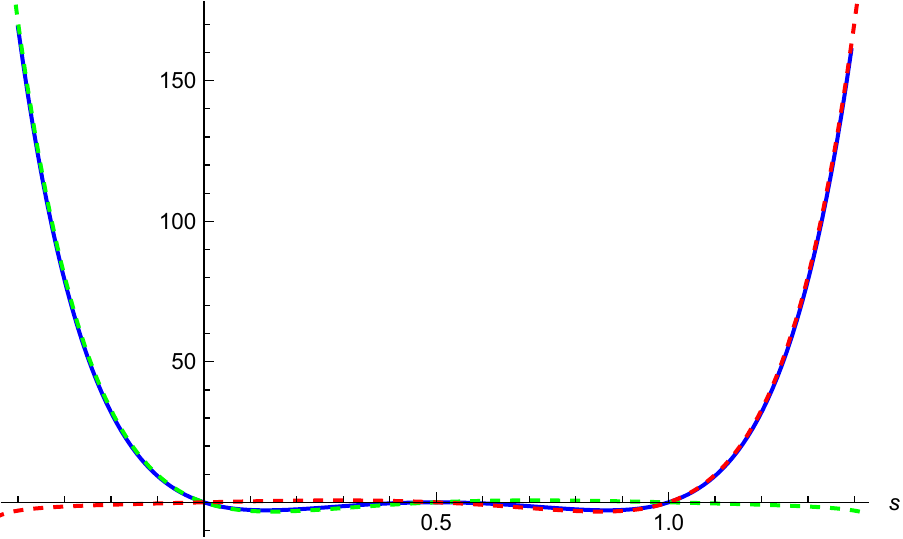}
  \caption{
Plot of  $c(s,N)$. 
The blue solid curve is the exact result of $c(s,N)$ while
the red and green dashed curves represent $p(s) N^{1+s}$ and $p(1-s) N^{2-s}$, respectively.
In this figure we set $N=15$.
}
  \label{fig:cs}
\end{figure}

In this paper, we are rather interested in the large $N$ limit. Since we already have the exact form for finite $N$, we can see its large $N$ behavior numerically. It turns out that $c(s,N)$ has different leading large $N$ behaviors for $s>1/2$
and $s<1/2$ (see Figure~\ref{fig:cs}).
\begin{equation}
c(s,N)\approx\left\{
\begin{aligned}
 &p(s) N^{1+s},\quad &(s>1/2),\\
 &p(1-s) N^{2-s},\quad &(s<1/2),
\end{aligned} \right.
\end{equation}
where
\begin{equation}
\begin{aligned}
p(s):=\frac{4^{s-1}(s-1)(2s-1)\Gamma(s+1/2)}{\sqrt{\pi}\Gamma(s+2)\Gamma(s)}.
\end{aligned}
\end{equation}

Recall that $c(s,N)$ is exactly symmetric under $s \leftrightarrow 1-s$. This symmetry suggests that the large $N$ asymptotic expansion of $c(s,N)$ has the following symmetric form:\footnote{This large $N$ expansion is slightly different from the one in \cite{Dorigoni:2021guq}, in which the authors seemed to assume $s>1/2$ implicitly.}
\begin{equation}
\begin{aligned}
c(s,N)=p(s) N^{1+s} \sum_{g=0}^\infty \frac{f_g(s)}{N^{2g}}+p(1-s) N^{2-s} \sum_{g=0}^\infty \frac{f_g(1-s)}{N^{2g}}.
\end{aligned}
\label{eq:c-large-N}
\end{equation}
We will check the validity of this conjectural expansion later.
The first few terms of $f_g(s)$ were obtained in \cite{Dorigoni:2021guq},
\begin{equation}
\begin{aligned}
f_0(s)&=1, \\
f_1(s)&=\frac{(s+1)s(s-1)(s-6)}{24(2s-3)}, \\
f_2(s)&=\frac{(s+1)s(s-1)(s-2)(s-3)(s-4)(5s^2-47s+30)}{5760(2s-3)(2s-5)}.
\end{aligned}
\end{equation}
Here we show a recursion relation to determine $f_g(s)$ very efficiently.
The key relation is the following difference equation for $c(s,N)$:
\begin{equation}
\begin{aligned}
N(N-1)c(s,N+1)-\Bigl[2(N^2-1)+s(s-1)\Bigr]c(s,N)+N(N+1)c(s,N-1)=0,
\end{aligned}
\label{eq:c-difference}
\end{equation}
as argued in \cite{Dorigoni:2021guq}. Since this is a second order difference equation for $N$, it has two independent solutions. This equation is also invariant under $s \leftrightarrow 1-s$.  These properties imply that the two asymptotic series on the right hand side in \eqref{eq:c-large-N} correspond to two independent solutions to the difference equation  \eqref{eq:c-difference}.

Let us introduce $p(s,N)$ as an asymptotic expansion:
\begin{equation}
\begin{aligned}
p(s,N):=p(s) N^{1+s} \sum_{g=0}^\infty \frac{f_g(s)}{N^{2g}}.
\end{aligned}
\label{eq:hat-p}
\end{equation}
Then, the large $N$ expansion \eqref{eq:c-large-N} is simply written as
\begin{equation}
\begin{aligned}
c(s,N)=p(s,N)+p(1-s,N).
\end{aligned}
\label{eq:c-pp}
\end{equation}
As discussed above, $p(s,N)$ should be a solution to \eqref{eq:c-difference}.
Therefore, we substitute the asymptotic series \eqref{eq:hat-p} into the difference equation \eqref{eq:c-difference}. 
To do so, we rewrite $p(s,N\pm 1)$ as
\begin{equation}
\begin{aligned}
p(s,N\pm 1)&=p(s) \sum_{g=0}^\infty f_g(s)(N\pm 1)^{1+s-2g} \\
&=p(s)N^{1+s} \sum_{g=0}^\infty \frac{f_g(s)}{N^{2g}} \sum_{j=0}^\infty \binom{1+s-2g}{j} \left( \pm \frac{1}{N} \right)^j .\\
\end{aligned}
\end{equation}
In the computation, we need the sum/difference of them, and we have more convenient forms
\begin{equation}
\begin{aligned}
p(s,N+1)+p(s,N-1)&=2p(s)N^{1+s} \sum_{g=0}^\infty N^{-2g} \sum_{h=0}^g \binom{1+s-2h}{2g-2h} f_h(s), \\
p(s,N+1)-p(s,N-1)&=2p(s)N^{s} \sum_{g=0}^\infty N^{-2g} \sum_{h=0}^g \binom{1+s-2h}{1+2g-2h} f_h(s).
\end{aligned}
\end{equation}
After some lengthy computations, we find the following recursion relation:
\begin{equation}
\begin{aligned}
f_g(s)=\sum_{h=0}^{g-1}\frac{(2+4g-2h-s)\Ga(2+s-2h)}{g(1+2g-2s)\Ga(3+2g-2h)\Ga(1+s-2g)}f_h(s).
\end{aligned}
\end{equation}
This equation determines $f_g(s)$ from $f_h(s)$ ($0\leq h \leq g-1$). Therefore, starting with $f_0(s)=1$, all the functions $f_g(s)$ are uniquely fixed. One can quickly compute $f_g(s)$ up to very high genera.

\subsection{Pole cancellation}
We show a simple consistency test for the large $N$ expansion \eqref{eq:c-large-N}. 
The function $c(s,N)$ is defined for arbitrary $s$. However, the coefficients $f_g(s)$ have poles at $s=3/2, 5/2, \dots$, and $p(s)$ has poles at $s=-1/2, -3/2, \dots$. This means that $p(s,N)$ itself is ill-defined for these values of $s$.
We observe that the combination $p(s,N)+p(1-s,N)$ is well-defined for any $s$. All the singular terms in $p(s,N)$ are canceled by those in $p(1-s,N)$!
As an example, let us check this property at $s=3/2$.
At order $\cO(N^{1/2})$, we find the following singular behaviors:
\begin{equation}
\begin{aligned}
 \lim_{s\to3/2}p(s)N^{1+s} \frac{f_1(s)}{N^2}&=-\frac{3\rt{N}}{8\pi^{3/2}(s-3/2)}+\cdots ,\\
\lim_{s\to3/2}p(1-s)N^{2-s} \frac{f_0(1-s)}{N^0}&=+\frac{3\rt{N}}{8\pi^{3/2}(s-3/2)}+\cdots .
\end{aligned} 
\end{equation}
Thus the poles at $s=3/2$  cancel to each other at this order. 
In general we expect that the pole at $s=k+1/2~(k=1,2,\cdots)$ is canceled
order by order in the genus expansion
\begin{equation}
\begin{aligned}
\lim_{s\to k+\hf}\Bigl[p(s)N^{1+s} \frac{f_{g+k}(s)}{N^{2(g+k)}}+
p(1-s)N^{2-s} \frac{f_{g}(1-s)}{N^{2g}} \Bigr]=(\text{regular}).
\end{aligned} 
\label{eq:pole-fg}
\end{equation} 
We have checked this relation for various $k$ and $g$.
It would be interesting to find a general proof of this relation.

\subsection{Borel resummation}
The systematic computation of $f_g(s)$ allows us to resum the large $N$ expansion \eqref{eq:c-large-N}.
Here we numerically show that the large $N$ resummation correctly reproduces the exact result for finite $N$.
We use the so-called Borel-Pad\'e resummation. Its main advantage is that one can apply it to factorially divergent series.

Let us briefly review the Borel-Pad\'e resummation technique. 
Let us consider a formal $1/N$ expansion
\begin{equation}
\begin{aligned}
F(N)=\sum_{g=0}^\infty \frac{F_g}{N^{2g}}.
\end{aligned}
\label{eq:F-formal}
\end{equation}
We first define the Borel transform by
\begin{equation}
\begin{aligned}
\cB_a[F](\zeta):=\sum_{g=0}^\infty \frac{F_g}{\Gamma(2g+a)}\zeta^{2g},
\end{aligned}
\label{eq:B-F}
\end{equation}
with a parameter $a$.
Then, the Borel resummation of \eqref{eq:F-formal} is given by
\begin{equation}
\begin{aligned}
F_a^\text{Borel}(N)=\int_0^\infty d\zeta\, e^{-\zeta} \zeta^{a-1} \cB_a[F] \left( \frac{\zeta}{N} \right)
\end{aligned}
\end{equation}
Typically, the parameter $a$ is set as unity, but we can choose it whatever 
as we want.
In practice, we have $F_g$ up to a finite order. In this case, we replace the Borel transform \eqref{eq:B-F} by its Pad\'e approximant.
We refer to this procedure as the Borel-Pad\'e resummation.

We apply this procedure to $p(s,N)$.
We repeat the computation for various $s$ and $N$. The result is shown in Table~\ref{tab:Borel-resum}.
We confirmed that the naive Borel-Pad\'e resummation of the combination $p(s,N)+p(1-s,N)$ correctly reproduces the exact value of $c(s, N)$.
This means that there are no non-perturbative corrections to $c(s, N)$ 
in the large $N$ limit.
On the other hand, one can also consider the large $N$ expansion of 
$\cG_N^{(\text{0-inst})}(M_{1,0})$. We find that this large $N$ expansion receives non-perturbative corrections. This difference comes from 
an exchange of the summation and the integration, 
as we will see in the next section.

\begin{table}[tp]
\caption{The Borel-Pad\'e resummation of the asymptotic $1/N$ expansion \eqref{eq:c-large-N}. The individual resummation $p^\text{Borel}(s,N)$ or $p^\text{Borel}(1-s,N)$ does not reproduce the exact result of $c(s,N)$. The sum of them does. We used $f_g(s)$ up to $g=120$ and the diagonal Pad\'e approximant of the Borel transform. The numerical values of the large $N$ resummation show a remarkable agreement even for $N=2$. We also set $a=1$ in the numerical computation.}
\begin{center}
\begin{tabular}{ccccc}
\hline
$s$ & $N$ & $p^\text{Borel}(s,N)$ &  $p^\text{Borel}(1-s,N)$ & $\displaystyle \left|1-\frac{p^\text{Borel}(s,N)+p^\text{Borel}(1-s,N)}{c(s,N)}\right|$ \\
\hline
$2/3$ & $2$ & $-0.0563855378210$ & $0.0440398588087$ & $6.32 \times 10^{-46}$ \\
	  & $3$ & $-0.111597704486$ & $0.0759324095617$ & $1.00\times 10^{-56}$ \\
	  & $4$ & $-0.180689554674$ & $0.111594396415$ & $1.05\times 10^{-65}$ \\
\hline
$5/4$ & $2$ & $0.605741506110$ & $-0.254179006110$ & $6.78 \times 10^{-45}$ \\
	& $3$ & $1.45334429406$ & $-0.343725153435$ & $1.02 \times 10^{-55}$ \\
	& $4$ & $2.73979678726$ & $-0.426144748682$ & $1.03 \times 10^{-64}$ \\
\hline                     
$\sqrt{3}$ & $2$ & $3.52020101650$ & $0.329163888886$ & $3.51 \times 10^{-44}$ \\
		& $3$ & $13.6225376249$ & $0.365956652856$ & $4.91 \times 10^{-55}$ \\
		& $4$ & $32.1651458968$ & $0.394892370253$ & $4.68 \times 10^{-64}$ \\
\hline
\end{tabular}
\end{center}
\label{tab:Borel-resum}
\end{table}%


\section{Worldsheet instanton and D-string instanton}\label{sec:WS-D1-inst}
As we have shown in \eqref{eq:pq-sum}, the full integrated correlator $\cG_N$ is constructed by the single function $\cG_N^{(\text{0-inst})}(M)$, 
the zero instanton part. In this section, we consider the large $N$ expansion of 
the F-string contribution $\cG_N^{(\text{0-inst})}(M_{1,0})$. 
It turns out that there are non-perturbative corrections to this expansion.

\subsection{Genus expansion}
Let us derive the genus expansion of 
$\cG_N^{(\text{0-inst})}(M_{1,0})$.
We start with the following integral representation 
of the full correlator $\cG_N$ \cite{Collier:2022emf}:
\begin{equation}
\begin{aligned}
 \cG_N&=\frac{N(N-1)}{4}-\hf\int_{\frac{1}{2}-\ri\infty}^{\frac{1}{2}+\ri\infty} \frac{ds}{2\pi\ri}\frac{\pi }{\sin\pi s}c(s,N)E_s^*\\
\end{aligned} 
\label{eq:G_N-Barnes}
\end{equation}
where $E_s^*$ denotes the non-holomorphic Eisenstein series
\begin{equation}
\begin{aligned}
 E_s^*=\Ga(s)\sum_{(m,n)\ne(0,0)}(M_{m,n})^{-s}
=\int_0^\infty dt\,t^{s-1}\Bigl[\Th(t,\tau)-1\Bigr].
\end{aligned} 
\end{equation}
To proceed, we further use the Fourier expansion of $E_s^*$, 
\begin{equation}
\begin{aligned}
 E_s^*=\sum_{k\in\mZ}e^{\ri k\th}\cF_k
\end{aligned} 
\end{equation}
where
\begin{equation}
\cF_k=\left\{
\begin{aligned}
 &2\Ga(s)\zeta(2s)(M_{1,0})^{-s}+2\Ga(1-s)\zeta(2-2s)(M_{1,0})^{s-1},&(k=0),\\
&4|k|^{s-\hf}\si_{1-2s}(|k|)
\rt{\tau_2}K_{s-\hf}(2\pi |k|\tau_2),
&(k\ne0).
\end{aligned}\right. 
\label{eq:F_k}
\end{equation}
Above, $K_\nu(x)$ denotes the modified Bessel function of
the second kind and $\si_s(n)$ is the divisor function
\begin{equation}
\begin{aligned}
 \si_s(n)=\sum_{d|n}d^s.
\end{aligned} 
\end{equation}
See e.g. \cite{Dorigoni:2021guq} for a derivation of \eqref{eq:F_k}.

Our interest is the zero-instanton sector, which is originally given by \eqref{eq:0-inst}.
With the help of the integral representation \eqref{eq:G_N-Barnes}, we also obtain another integral representation,
\begin{equation}
\begin{aligned}
 \cG_N^{(\text{0-inst})}(M_{1,0})&=\frac{N(N-1)}{4}-\frac{1}{2} \int_{\frac{1}{2}-\ri\infty}^{\frac{1}{2}+\ri\infty} \frac{ds}{2\pi\ri}\frac{\pi }{\sin\pi s}c(s,N) \cF_0.
\end{aligned} 
\end{equation}
Plugging \eqref{eq:c-pp} and \eqref{eq:F_k} into this integral, we have four contributions.
Since there are simple symmetric relations
\begin{equation}
\begin{aligned}
&\int_{\frac{1}{2}-\ri\infty}^{\frac{1}{2}+\ri\infty} \frac{ds}{2\pi\ri}\frac{\pi }{\sin\pi s}p(1-s,N)\Gamma(1-s)\zeta(2-2s)(M_{1,0})^{s-1}\\
&\qquad=\int_{\frac{1}{2}-\ri\infty}^{\frac{1}{2}+\ri\infty} \frac{ds}{2\pi\ri}\frac{\pi }{\sin\pi s}p(s,N)\Gamma(s)\zeta(2s)(M_{1,0})^{-s},\\
&\int_{\frac{1}{2}-\ri\infty}^{\frac{1}{2}+\ri\infty} \frac{ds}{2\pi\ri}\frac{\pi }{\sin\pi s}p(s,N)\Gamma(1-s)\zeta(2-2s)(M_{1,0})^{s-1}\\
&\qquad=\int_{\frac{1}{2}-\ri\infty}^{\frac{1}{2}+\ri\infty} \frac{ds}{2\pi\ri}\frac{\pi }{\sin\pi s}p(1-s,N)\Gamma(s)\zeta(2s)(M_{1,0})^{-s},
\end{aligned}
\end{equation}
we obtain
\begin{equation}
\begin{aligned}
 \cG_N^{(\text{0-inst})}(M_{1,0})=\frac{N(N-1)}{4}+I_N+\til{I}_N,
\end{aligned}
\label{eq:0-inst-int}
\end{equation}
where
\begin{equation}
\begin{aligned}
I_N&:= -\int_{\frac{1}{2}-\ri\infty}^{\frac{1}{2}+\ri\infty} \frac{ds}{2\pi\ri}\frac{2\pi }{\sin\pi s}p(s,N)\Gamma(1-s)\zeta(2-2s)(M_{1,0})^{s-1}, \\
\til{I}_N&:= -\int_{\frac{1}{2}-\ri\infty}^{\frac{1}{2}+\ri\infty} \frac{ds}{2\pi\ri}\frac{2\pi }{\sin\pi s}p(1-s,N)\Gamma(1-s)\zeta(2-2s)(M_{1,0})^{s-1}.
\end{aligned}
\end{equation}
Using the large $N$ expansion \eqref{eq:hat-p} and exchanging the sum and the integral, these functions admit the genus expansions,
\begin{equation}
\begin{aligned}
I_N&=\sum_{g=0}^\infty N^{2-2g} \biggl[ -\int_{\frac{1}{2}-\ri\infty}^{\frac{1}{2}+\ri\infty} \frac{ds}{2\pi\ri}\frac{2\pi }{\sin\pi s}p(s)f_g(s)\Gamma(1-s)\zeta(2-2s) ( M_{1,0}N )^{s-1} \biggr], \\
\til{I}_N&=\sum_{g=0}^\infty N^{1-2g} \biggl[ -\int_{\frac{1}{2}-\ri\infty}^{\frac{1}{2}+\ri\infty} \frac{ds}{2\pi\ri}\frac{2\pi }{\sin\pi s}p(1-s)f_g(1-s)\Gamma(1-s)\zeta(2-2s) ( M_{1,0}^{-1}N )^{1-s} \biggr],
\end{aligned}
\label{eq:GN-tGN}
\end{equation}
where $M_{1,0}N$ and $M_{1,0}^{-1}N$ are related to the 't Hooft coupling $\la$
and the dual 't Hooft coupling $\til{\la}$ by
\begin{equation}
\begin{aligned}
 \la&=4\pi g_sN=4M_{1,0}N,\\
\til{\la}&=\frac{\la}{g_s^2}=4\pi g_s^{-1}N=4\pi^2 M_{1,0}^{-1}N. 
\end{aligned} 
\end{equation}
We should emphasize that the dual 't Hooft coupling  $\til{\la}$ is not the S-dual
of $\la$
\begin{equation}
\begin{aligned}
 \til{\la}\ne 4M_{0,1}N.
\end{aligned} 
\end{equation}

This is not the end of the story. Each of these functions turns out to give unusual genus expansions. 
For instance, in the weak coupling regime $\lambda \ll 1$, one can close the integration contour in $I_N$ by adding the infinite right semi-circle.
Then one obtains the following weak coupling expansion of $I_N$,
\begin{equation}
\begin{aligned}
I_N&=N^2 \biggl(-\frac{1}{4}+\frac{3\zeta(3)}{8\pi^2}\lambda-\frac{75\zeta(5)}{128\pi^4}\lambda^2+\cO(\lambda^3) \biggr) \\
&\quad+N^0 \biggl(\frac{\lambda^{1/2}}{16}-\frac{3\zeta(3)}{8\pi^2}\lambda+\frac{75\zeta(5)}{128\pi^4}\lambda^2+\cO(\lambda^3) \biggr)\\
&\quad+N^{-2} \biggl(\frac{13\lambda^{1/2}}{8192}-\frac{\lambda^{3/2}}{1536}+\frac{945\zeta(9)}{4096\pi^8}\lambda^4+\cO(\lambda^5) \biggr)+\cO(N^{-4}).
\end{aligned}
\label{eq:IN-expand}
\end{equation}
The weak coupling expansion has unusual fractional power terms.
Such terms are however canceled by taking into account the counterpart $\til{I}_N$, as discussed in \cite{Collier:2022emf}. 
To see it, we rewrite $\til{I}_N$ as
\begin{equation}
\begin{aligned}
\til{I}_N=\sum_{g=0}^\infty N^{1-2g} \biggl[ -\int_{\frac{1}{2}-\ri\infty}^{\frac{1}{2}+\ri\infty} \frac{ds}{2\pi\ri}\frac{2\pi }{\sin\pi s}p(1-s)f_g(1-s)\Gamma(1-s)\zeta(2-2s) \biggl( \frac{\la}{4N} \biggr)^{s-1} \biggr]
\end{aligned}
\end{equation}
Since in the 't Hooft limit, we first take $N \to \infty$ by keeping $\la$ finite, $\la/(4N)$ is less than unity for any $\la$ in this limit.
Therefore, 
we should close the integration contour of $\til{I}_N$
in the same way as $I_N$
by adding the infinite right semi-circle.
Then we find
\begin{equation}
\begin{aligned}
\til{I}_N&=\frac{N}{4}-N^0\frac{\lambda^{1/2}}{16}+N^{-2}\biggl(-\frac{13\lambda^{1/2}}{8192}+\frac{\lambda^{3/2}}{1536} \biggr)+\cO(N^{-4}),
\end{aligned}
\end{equation}
which cancels the fractional power terms of $\la$ in \eqref{eq:IN-expand}.
Therefore we conclude that the combination $I_N+\til{I}_N$ gives the 
usual genus expansion without the fractional power terms of $\la$.%
\footnote{In \cite{Collier:2022emf}, the condition \eqref{eq:pole-fg}
was obtained by requiring the absence of the fractional power terms of $\la$
in the small $\la$ expansion.
We should stress that our logic of the derivation of \eqref{eq:pole-fg}
is different from \cite{Collier:2022emf}:
we derived the condition \eqref{eq:pole-fg} from the regularity of the combination
$p(s,N)+p(1-s,N)$ in \eqref{eq:c-pp} 
and the absence of the fractional power terms of $\la$
is a consequence of \eqref{eq:pole-fg}, not the other way around.}
Let us denote
\begin{equation}
\begin{aligned}
\cG_N^{\text{(0-inst)}}(M_{1,0})\sim \sum_{g=0}^\infty N^{2-2g} G_g(\lambda),
\end{aligned}
\label{eq:0-inst-genus}
\end{equation}
where
\begin{equation}
\begin{aligned}
 G_g(\la):=
-\int_{c_g-\ri\infty}^{c_g+\ri\infty}\frac{ds}{2\pi\ri}\frac{2\pi }{\sin\pi s}p(s)f_g(s)
\Ga(1-s)\zeta(2-2s)\biggl(\frac{\la}{4}\biggr)^{s-1}.
\end{aligned} 
\label{eq:GtilG-sint}
\end{equation}
We have used $\sim$ in \eqref{eq:0-inst-genus} because these genus expansions receive non-perturbative corrections, as we will see later. Note also that the integration contour in this expression should be chosen so that the weak coupling expansion in $\la$ does not include fractional power terms. It is not hard to find them. For $g=0,1$, we find
\begin{equation}
\begin{aligned}
1 < c_0 < 2,\qquad \frac{3}{2} < c_1 < 2.
\end{aligned}
\end{equation} 
For $g \geq 2$, we have
\begin{equation}
\begin{aligned}
g+\frac{1}{2} < c_g < 2g+1 \qquad (g \geq 2).
\end{aligned}
\end{equation}

We can also derive another genus expansion in terms of the dual coupling $\til{\la}$.
For $\til{\la} \ll 1$, $\til{I}_N$ admits the following weak coupling expansion:
\begin{equation}
\begin{aligned}
\til{I}_N&=N \biggl( \frac{\til{\la}^2}{480}-\frac{5\til{\la}^3}{12096}+\frac{7\til{\la}^4}{115200}+\cO(\til{\la}^5) \biggr) \\
&\quad+N^{-1} \biggl( \frac{3\zeta(3)\til{\la}^{3/2}}{64\pi^3}-\frac{\til{\la}^2}{480}+\frac{5\til{\la}^3}{12096}+\cO(\til{\la}^4) \biggr) \\
&\quad+N^{-3}\biggl( \frac{39\zeta(3)\til{\la}^{3/2}}{32768\pi^3}-\frac{45\zeta(5)\til{\la}^{5/2}}{4096\pi^5}+\frac{\til{\la}^5}{443520}+\cO(\til{\la}^6) \biggr)+\cO(N^{-5}).
\end{aligned}
\end{equation}
The fractional powers are canceled by $I_N$:
\begin{equation}
\begin{aligned}
I_N=-N^{-1}\frac{3\zeta(3)\til{\la}^{3/2}}{64\pi^3}+N^{-3}\biggl(-\frac{39\zeta(3)\til{\la}^{3/2}}{32768\pi^3}+\frac{45\zeta(5)\til{\la}^{5/2}}{4096\pi^5}\biggr)+\cO(N^{-5}).
\end{aligned}
\end{equation}
Therefore in this case, we obtain
\begin{equation}
\begin{aligned}
\cG_N^{\text{(0-inst)}}(M_{1,0})\sim \frac{N(N-1)}{4}+ \sum_{g=0}^\infty N^{1-2g} \til{G}_g(\til{\la}),
\end{aligned}
\end{equation}
where
\begin{equation}
\begin{aligned}
\til{G}_g(\til{\la}):=
-\int_{\til{c}_g-\ri\infty}^{\til{c}_g+\ri\infty}\frac{ds}{2\pi\ri}\frac{2\pi }{\sin\pi s}p(1-s)f_g(1-s)
\Ga(1-s)\zeta(2-2s)\biggl(\frac{\til{\la}}{4\pi^2}\biggr)^{1-s}.
\end{aligned}
\end{equation}
The real number $\til{c}_g$ should be chosen so that the weak coupling expansion in $\til{\la}$ does not include any fractional powers. It is easy to check that we can set $\til{c}_g=1-c_g$.
Therefore the same result is obtained by using the same contour in $G_g(\la)$,
\begin{equation}
\begin{aligned}
\til{G}_g(\til{\la})=-\int_{c_g-\ri\infty}^{c_g+\ri\infty}\frac{ds}{2\pi\ri}\frac{2\pi }{\sin\pi s}p(s)f_g(s)
\Ga(s)\zeta(2s)\biggl(\frac{\til{\la}}{4\pi^2}\biggr)^{s}.
\end{aligned}
\label{eq:tilG_g}
\end{equation}

\subsection{Worldsheet instanton}
After identifying the integration contours, we can discuss the strong coupling expansions.
Let us first consider $G_g(\la)$.
Our logic in this subsection is as follows. The large $\la$ expansion of $G_g(\la)$ is Borel non-summable.
This means that the Borel resummation of the large $\la$ expansion has a discontinuity due to an ambiguity of the resummation.
It is well-known that this discontinuity is related to non-perturbative corrections, which are interpreted as worldsheet instanton corrections in the present case.

The large $\la$ expansion is obtained by closing the contour along the infinite left semi-circle.
In the following, we will often omit the contour of the $s$-integrals in $G_g(\la)$ and $\til{G}_g(\til{\la})$ if it is the path from $c_g-\ri\infty$ to $c_g+\ri\infty$.
We first rewrite $G_g(\la)$ in \eqref{eq:GtilG-sint} as
\begin{equation}
\begin{aligned}
 G_g(\la)&=
-\int\frac{ds}{2\pi\ri}\frac{\pi }{\sin\pi s}p(s)f_g(s)
\Ga(1-s)\sum_{n=1}^\infty2(n^2M_{1,0}N)^{s-1}\\
&=
-\sum_{n=1}^\infty 2nT_F\int_0^\infty 
 dw e^{-nA_{F}w}\int\frac{ds}{2\pi\ri}\frac{\pi }{\sin\pi s}p(s)f_g(s)\Ga(1-s)
\frac{(4w)^{2-2s}}{\Ga(3-2s)}
\end{aligned} 
\end{equation}
where we used $\zeta(2-2s)=\sum_{n=1}^\infty n^{2s-2}$ and $T_F$ is defined by
\begin{equation}
\begin{aligned}
 T_F=4\rt{M_{1,0}N}=2\rt{\la}.
\end{aligned} 
\end{equation}
At strong coupling, by picking up the poles, this can be written as
\begin{equation}
\begin{aligned}
 G_g(\la)&=\sum_{n=1}^\infty 2nT_F\int_0^\infty 
 dw\, e^{-nA_{F}w}[\phi_g^+(w)+\phi_g^-(w)],
\end{aligned} 
\label{eq:Gg-phig}
\end{equation}
where $\phi_g^+(w)$, $\phi_g^-(w)$ are residue contributions from poles on the right/left half $s$-plane, respectively. There are finite number of poles on the strip $0<\text{Re}\, s<c_g$, while there are infinite number of poles on the left half $s$-plane.

Since we are interested in the discontinuity, the negative part $\phi_g^-(w)$ is important, which is given by
\begin{equation}
\begin{aligned}
 \phi_g^-(w)=\sum_{\ell=0}^\infty \bt_{\ell}f_g(-\ell-1/2)
\frac{(4w)^{2\ell+3}}{\Ga(2\ell+4)}
\end{aligned} 
\label{eq:phi-g}
\end{equation}
with $\bt_\ell$ being
\begin{equation}
\begin{aligned}
 \bt_\ell=\text{Res}_{s=-\ell-\hf}\left[-\frac{\pi}{\sin\pi s}p(s)\Ga(1-s)\right]
=\frac{\Ga\left(\ell-\hf\right)\Ga\left(\ell+\frac{5}{2}\right)\Ga(2\ell+3)}{
2^{4\ell+4}\pi\Ga^2(\ell+1)}.
\end{aligned} 
\end{equation}
As we will see in the next section, the large order behavior of the genus sum is governed by the positive part $\phi_g^+(w)$.

For the genus-zero part we find the following analytic expression:
\begin{equation}
\begin{aligned}
 \phi_0^-(w)&=\sum_{\ell=0}^\infty \bt_{\ell}
\frac{(4w)^{2\ell+3}}{\Ga(2\ell+4)}
=-2w^3\, {}_2F_1\Bigl(-\frac{1}{2},\frac{3}{2},1,w^2\Bigr).
\end{aligned} 
\end{equation}
This agrees with the result in \cite{Dorigoni:2021guq}.
$\phi_0^-(w)$ has a branch cut along 
the positive real $w$-axis running from $w=1$ to $w=\infty$.
To see the discontinuity of $\phi_0^-(w)$ across the cut, it is convenient to rewrite it in terms of the complete elliptic integrals
\begin{equation}
\begin{aligned}
 \phi_0^-(w)=\frac{4w^3}{\pi}\Bigl[K(w^2)-2E(w^2)\Bigr],
\end{aligned} 
\end{equation}
where $K(w^2)$ and $E(w^2)$ denote the complete elliptic integral of the first and the second kind, respectively.
From the known discontinuity of $K(z)$ and $E(z)$
\begin{equation}
\begin{aligned}
 K(z+\ri 0)-K(z-\ri 0)&=2\ri K(1-z),\\
E(z+\ri 0)-E(z-\ri 0)&=2\ri K(1-z)-2\ri E(1-z),
\end{aligned} 
\label{eq:discKE}
\end{equation}
we find the discontinuity of $\phi_0^-(w)$
\begin{equation}
\begin{aligned}
 \text{Disc}\,\phi_0^-(w)&:=\phi_0^-(w+\ri 0)-\phi_0^-(w-\ri 0)\\
&=-\frac{8\ri w^3}{\pi}\Bigl[K(1-w^2)-2E(1-w^2)\Bigr]\\
&=4\ri w^3
\, {}_2F_1\Bigl(-\frac{1}{2},\frac{3}{2},1,1-w^2\Bigr).
\end{aligned} 
\end{equation}
This reproduces eq.(5.35) in \cite{Dorigoni:2021guq}.
Then the worldsheet instanton corrections to $G_0(\la)$ 
is computed by expanding $\text{Disc}\,\phi_0^-(w)$
around $w=1$ and integrate it term by term
\begin{equation}
\begin{aligned}
 \lap G_0&=8\ri\int_1^\infty dw\sum_{n=1}^\infty nT_F
e^{-nT_Fw} w^3
\, {}_2F_1\Bigl(-\frac{1}{2},\frac{3}{2},1,1-w^2\Bigr)\\
&=8\ri \sum_{n=1}^\infty e^{-nT_F}
\left[1+\frac{9}{2 nT_F}+\frac{117}{8(nT_F)^2}+\frac{489}{16(nT_F)^3}
+\cdots\right]\\
&=8\ri \left[\cL_0(T_F)+\frac{9}{2}\cL_1(T_F)
+\frac{117}{8}\cL_2(T_F)+
\frac{489}{16}\cL_3(T_F)
+\cdots\right],
\end{aligned} 
\label{eq:lap-G0}
\end{equation}
where we defined
\begin{equation}
\begin{aligned}
 \cL_m(T_F)=\sum_{n=1}^\infty e^{-nT_F}(nT_F)^{-m}=
(T_F)^{-m}\text{Li}_m(e^{-T_F}),
\end{aligned} 
\label{eq:polylog}
\end{equation}
and $\text{Li}_m(e^{-T_F})$ denotes the polylogarithm.
Note that $T_F$ corresponds to the action of the worldsheet one-instanton
and the sum over $n$ 
in \eqref{eq:polylog} is interpreted as the multi-covering of the worldsheet 
instantons.
The appearance of the polylog is reminiscent of the worldsheet
instanton corrections in the topological string. 

In general, $\lap G_0$ is expanded as 
\begin{equation}
\begin{aligned}
 \lap G_0&=8\ri\sum_{m=0}^\infty a_{m} \cL_m(T_F)=8\ri \sum_{n=1}^\infty e^{-n T_F} \sum_{m=0}^\infty \frac{a_{m}}{(nT_F)^m}.
\end{aligned} 
\label{eq:delta-G0}
\end{equation}
As discussed in \cite{Dorigoni:2021guq}, $a_{m}$ obeys the recursion relation
\begin{equation}
\begin{aligned}
 a_{m}&=-\frac{3 \left(2 m^4-16 m^3+31 m^2+13 m-15\right)
}{2m \left(2 m^2-10 m+3\right)} a_{m-1}\\
&\quad-\frac{(m-6) (m-2) \left(2 m^2-6 m-5\right)}{2 \left(2 m^2-10 m+3\right)} a_{m-2}.
\end{aligned} 
\end{equation}
By using this recursion relation and the Richardson extrapolation
(see appendix \ref{app:richardson}), we find the large $m$ behavior, 
\begin{equation}
\begin{aligned}
 a_{m}\sim\frac{12}{\pi}(-1)^m\Ga(m-3),\qquad(m\gg1).
\end{aligned} 
\label{eq:am-large}
\end{equation}
Since $a_{m}$ has an alternating sign at large $m$, 
the summation over $m$ in \eqref{eq:delta-G0}
is now Borel summable. 
This implies that the median resummation of the large $\la$ expansion of $G_0(\la)$ is a resurgent function \cite{Dorigoni:2021guq}.

Next consider the genus-one part.
For $g=1$ we find
\begin{equation}
\begin{aligned}
 \phi_1^-(w)&=\frac{\left(-41 w^4+81 w^2-36\right) K(w^2)}{96 \pi  w
   \left(w^2-1\right)^2}+\frac{\left(10 w^6-65 w^4+99 w^2-36\right)
   E(w^2)}{96 \pi  w \left(w^2-1\right)^3}\\
&=-\frac{1}{64w}\Biggl[
(10w^4+81w^2+36){}_2F_1\Bigl(\frac{3}{2},\frac{5}{2},1,w^2\Bigr)
+9(3w^4+w^2-4){}_2F_1\Bigl(\frac{5}{2},\frac{5}{2},1,w^2\Bigr)
\Biggr].
\end{aligned} 
\label{eq:phi-g1}
\end{equation}
Using \eqref{eq:discKE}, we find the discontinuity of $\phi_1^-(w)$
\begin{equation}
\begin{aligned}
&\text{Disc}\,\phi_1^-(w)\\
=&-\frac{\ri w \left(31 w^4-57 w^2+18\right) K(1-w^2)}{48 \pi 
   \left(w^2-1\right)^3}-\frac{\ri \left(10 w^6-65 w^4+99 w^2-36\right)
   E(1-w^2)}{24 \pi  w \left(w^2-1\right)^3}\\
=&-\frac{\ri}{2^{11}w}
\biggl[8(10w^4+81w^2+36){}_2F_1\Bigl(\frac{3}{2},\frac{5}{2},4,1-w^2\Bigr)
+27(3w^4+w^2-4){}_2F_1\Bigl(\frac{5}{2},\frac{5}{2},5,1-w^2\Bigr)
\biggr].
\end{aligned} 
\end{equation}
Then \eqref{eq:Gg-phig}
gives rise to the genus-one worldsheet instanton correction
\begin{equation}
\begin{aligned}
 &\sum_{n=1}^\infty 2nT_F\int_1^\infty dw\, e^{-nT_Fw}\text{Disc}\,\phi_1^-(w)\\
=&
\sum_{n=1}^\infty\ri e^{-nT_F}
\left[-\frac{127}{256}+\frac{927}{2048}(nT_F)^{-1}
-\frac{3897}{4096}(nT_F)^{-2}+\cdots\right]\\
=&\ri\left[-\frac{127}{256}\cL_0(T_F)
+\frac{927}{2048}\cL_1(T_F)
-\frac{3897}{4096}\cL_2(T_F)
+\cdots\right],
\end{aligned} 
\label{eq:G1-phidisc}
\end{equation} 
which agrees with eq.(5.45)  in \cite{Dorigoni:2021guq}.
However, it turns out that this is not the end of the story.
Since $\phi_1^-(w)$ has a pole at $w=1$, there are 
additional contributions to the imaginary part of $G_1(\la)$.
The Laurent expansion of $\phi_1^-(w)$ around $w=1$ reads
\begin{equation}
\begin{aligned}
  \phi_1^-(w)=\frac{1}{4\pi}\left[\frac{1}{24(w-1)^3}-\frac{3}{32(w-1)^2}-\frac{77}{128(w-1)}+\cO((w-1)^0)\right],
\end{aligned} 
\end{equation}
which also contributes to the imaginary part of $G_1(\la)$
\begin{equation}
\begin{aligned}
& -2\pi\ri\sum_{n=1}^\infty \text{Res}_{w=1}\Bigl[2nT_F e^{-nT_Fw}\phi_1^-(w)\Bigr] \\
=&\ri\sum_{n=1}^\infty e^{-nT_F}\left[-\frac{1}{48}(nT_F)^3-\frac{3}{32}(nT_F)^2
+\frac{77}{128}nT_F\right]\\
=& \ri\left[
-\frac{1}{48} \cL_{-3}(T_F)
-\frac{3}{32}\cL_{-2}(T_F)+\frac{77}{128}\cL_{-1}(T_F)\right].
\end{aligned} 
\label{eq:G1-res}
\end{equation}
Adding \eqref{eq:G1-phidisc} and \eqref{eq:G1-res}, the imaginary part of
$G_1(\la)$ becomes
\begin{equation}
\begin{aligned}
 \lap G_1=\ri\Biggl[&-\frac{1}{48} \cL_{-3}(T_F)
-\frac{3}{32}\cL_{-2}(T_F)+\frac{77}{128}\cL_{-1}(T_F)\\
&-\frac{127}{256}\cL_0(T_F)
+\frac{927}{2048}\cL_1(T_F)
-\frac{3897}{4096}\cL_2(T_F)+\cdots
\Biggr].
\end{aligned} 
\label{eq:lap-G1}
\end{equation}
We stress that the first three terms are missing in \cite{Dorigoni:2021guq}.
These terms should be included to re-construct the analytic function from the large $\la$ expansion of $G_1(\la)$.

In a similar manner, we find $\lap G_g$
for $g=2,3$
\begin{equation}
\begin{aligned}
 \lap G_2=\ri&\Biggl[\frac{1}{36864}\cL_{-6}(T_F)
+\frac{7}{40960}\cL_{-5}(T_F)
-\frac{857}{491520}\cL_{-4}(T_F)
-\frac{1681}{327680}\cL_{-3}(T_F)\\
&+\frac{28087}{2621440}\cL_{-2}(T_F)
+\frac{24825}{1048576}\cL_{-1}(T_F)
-\frac{153027}{4194304}\cL_{0}(T_F)
+\frac{726327}{8388608}\cL_{1}(T_F)+\cdots
\Biggr],\\
\lap G_3=\ri&\Biggl[-\frac{1}{42467328}\cL_{-9}(T_F)
-\frac{11}{47185920}\cL_{-8}(T_F)
+\frac{2489}{1321205760}\cL_{-7}(T_F)
+\frac{147661}{7927234560}\cL_{-6}(T_F)\\
&+\frac{94303}{2348810240}\cL_{-5}(T_F)
-\frac{3293543}{42278584320}\cL_{-4}(T_F)
-\frac{33749753}{56371445760}\cL_{-3}(T_F)
+\cdots
\Biggr].
\end{aligned} 
\label{eq:lap-G23}
\end{equation}
In general, the discontinuity of $G_g(\la)$ takes the form
\begin{equation}
\begin{aligned}
\lap G_g=\ri \sum_{m=0}^\infty a_{g,m}\cL_{-3g+m}(T_F)=\ri \sum_{n=1}^\infty e^{-n T_F} \sum_{m=0}^\infty a_{g,m}(nT_F)^{3g-m},
\end{aligned}
\end{equation}
where $a_{0,m}=8a_m$.

\subsection{D-string instanton}
As argued in \cite{Collier:2022emf}, the large $\til{\la}$
expansion of $\til{G}_g(\til{\la})$ is Borel non-summable and
hence there are non-perturbative ambiguity in the 
large $\til{\la}$
expansion of $\til{G}_g(\til{\la})$.
To see this, we rewrite  $\til{G}_g(\til{\la})$ in \eqref{eq:tilG_g}
into the multi-covering form
using the functional identity of the Riemann zeta function
\begin{equation}
\begin{aligned}
 \Ga(s)\zeta(2s)=\pi^{2s-\hf}\Ga\left(\hf-s\right)
\zeta(1-2s).
\end{aligned} 
\end{equation}
Then $\til{G}_g(\til{\la})$ in \eqref{eq:tilG_g}
is written as
\begin{equation}
\begin{aligned}
 \til{G}_g(\til{\la})&=
-\int\frac{ds}{2\pi\ri}\frac{\pi}{\sin\pi s}p(s)f_g(s)\sum_{n=1}^\infty
\frac{2}{n\rt{\pi}}\Ga(1/2-s)
(\pi^2 n^2M_{1,0}^{-1}N)^s\\
&=
-T_D\int\frac{ds}{2\pi\ri}\frac{\pi}{\sin\pi s}p(s)f_g(s)\sum_{n=1}^\infty
\int_0^\infty dw e^{-nT_Dw}(nT_D)^{1+3g}
\frac{w^{1+3g-2s}\Ga(1/2-s)}{2^{4s-1}\rt{\pi}\Ga(2+3g-2s)},
\end{aligned} 
\end{equation}
where
\begin{equation}
\begin{aligned}
 T_D=4\pi\rt{M_{1,0}^{-1}N}=2\rt{\til{\la}}=\frac{1}{g_s}T_F.
\end{aligned} 
\end{equation}
Closing the contour along the infinite left semi-circle and picking up the residues, we find
\begin{equation}
\begin{aligned}
 \til{G}_g(\til{\la})&=
T_D\sum_{n=1}^\infty \int_0^\infty dw e^{-nT_Dw}(nT_D)^{1+3g}
[\psi_g^+(w)+\psi_g^-(w)],
\end{aligned} 
\end{equation}
where $\psi_g^+(w)$, $\psi_g^-(w)$ are the contributions of poles from the right/left 
half $s$-plane.
In particular the negative pole part is
\begin{equation}
\begin{aligned}
 \psi_g^-(w)=\sum_{\ell=0}^\infty \bt_\ell
f_g(-\ell-1/2)
\frac{2^{4\ell+3}w^{2+3g+2\ell}\Ga(\ell+1)}{\rt{\pi}\Ga(\ell+3/2)\Ga(3+3g+2\ell)}.
\end{aligned} 
\label{eq:psi-g}
\end{equation}
The first few terms of $\psi_g^-(w)$ read
\begin{equation}
\begin{aligned}
 \psi_0^-(w)&=\frac{w^2(4w^2-3)}{2\pi\rt{1-w^2}},\\
\psi_1^-(w)&=\frac{1}{768\pi}\biggl[-\frac{w}{\rt{1-w^2}}-24\arcsin w+25w-18
\log\frac{1+\rt{1-w^2}}{2}\biggr],\\
\psi_2^-(w)&=\frac{1}{23592960\pi}
\biggl[-45 \left(13 w^2+120\right) w^2 \log\frac{1+\rt{1-w^2}}{2}
-\frac{5047 w^4-4501 w^2-586}{\rt{1-w^2}}\\
&\quad\quad+\frac{4907 w^4}{4}-4794 w^2-586\biggr].
\end{aligned} 
\label{eq:psi012}
\end{equation}
Some comments are in order here. In the definition of 
$\psi_g^-(w)$ in \eqref{eq:psi-g}, we have divided the summand 
by the factor
$\Ga(3+3g+2\ell)$ which depends on $g$.
Instead, we could have divided by the $g$-independent factor, e.g. 
$\Ga(3+2\ell)$. However, in that case 
$\psi_g^-(w)$ would have the singularity at $w=1$
of the form $(1-w^2)^{-m-\hf}$ with $m\geq1$.
In order to avoid such higher order singularity at $w=1$, we have divided 
in \eqref{eq:psi-g} by the factor
$\Ga(3+3g+2\ell)$. Then, as we can see from \eqref{eq:psi012},
the singularity of $\psi_g^-(w)$ at $w=1$ is at most $(1-w^2)^{-\hf}$
whose integral is convergent near $w=1$.\footnote{Similarly, if we replace 
$\Ga(2\ell+4)$ by $\Ga(2\ell+4+3g)$ 
in the definition of $\phi_g^-(w)$ in \eqref{eq:phi-g},
$\phi_g^-(w)$ does not have a pole at $w=1$.
Then we do not have to treat the contribution of the residue
at $w=1$ separately, as we did in \eqref{eq:G1-res}.
}

Taking the discontinuity of $\psi_g^-(w)$
across the cut running from $w=1$ to $w=\infty$, we find
the imaginary part of $\til{G}_g(\til{\la})$
\begin{equation}
\begin{aligned}
 \lap\til{G}_0&=\frac{\ri T_D}{\rt{2\pi}}
\biggl[\cL_{-\hf}(T_D)+\frac{39}{8}\cL_{\hf}(T_D)
+\frac{1785}{128}\cL_{\frac{3}{2}}(T_D)+
\frac{22365}{1024}\cL_{\frac{5}{2}}(T_D)
\cdots\biggr],\\
\lap\til{G}_1&=\frac{\ri T_D}{\rt{2\pi}}
\biggl[-\frac{1}{384}\cL_{-\frac{7}{2}}(T_D)
-\frac{17}{1024}\cL_{-\frac{5}{2}}(T_D)
+\frac{805}{16384}\cL_{-\frac{3}{2}}(T_D)\\
&\hspace{7truecm}
-\frac{4019}{131072}\cL_{-\hf}(T_D)
+\cdots\biggr],\\
\lap\til{G}_2&=\frac{\ri T_D}{\rt{2\pi}}
\biggl[\frac{1}{294912}\cL_{-\frac{13}{2}}(T_D)
+\frac{43}{1310720}\cL_{-\frac{11}{2}}(T_D) \\
&\hspace{3truecm}-\frac{2251}{20971520}\cL_{-\frac{9}{2}}(T_D)
-\frac{25099}{33554432}\cL_{-\frac{7}{2}}(T_D)
+\cdots\biggr].
\end{aligned} 
\label{eq:lap-tilG}
\end{equation}
In general, we have
\begin{equation}
\begin{aligned}
\lap \til{G}_g &=\frac{\ri T_D}{\rt{2\pi}} \sum_{m=0}^\infty \til{a}_{g,m} \cL_{-1/2-3g+m}(T_D)\\
&=\frac{\ri T_D}{\rt{2\pi}}\sum_{n=1}^\infty e^{-nT_D} \sum_{m=0}^\infty \til{a}_{g,m} (nT_D)^{1/2+3g-m}.
\end{aligned}
\label{eq:tilGg}
\end{equation}

Note that the discontinuity of $\psi_0^-(w)$ has a simple form
\begin{equation}
\begin{aligned}
 \text{Disc}\,\psi_0^-(w)=2\ri\frac{w^2(4w^2-3)}{2\pi\rt{w^2-1}},
\end{aligned} 
\end{equation}
and hence we can write down the expansion coefficients
of $\lap \til{G}_0$ in a closed form
\begin{equation}
\begin{aligned}
 \til{a}_{0,m}=\frac{(-1)^m 2^{-m-2} (4 m^2+35) \Gamma \left(m-\frac{7}{2}\right) \Gamma
   \left(m+\frac{5}{2}\right)}{\pi  \Gamma (m+1)}.
\end{aligned} 
\end{equation}
Since $\til{a}_{0,m}$ has an alternating sign, 
the summation over $m$ in \eqref{eq:tilGg} is Borel summable.
This is similar to the behavior of $a_{m}$ in \eqref{eq:am-large}.

In \cite{Collier:2022emf}, the imaginary part $\lap\til{G}_g$ is interpreted as the
contribution of D-string instanton.
However, we should stress that $\lap\til{G}_g$ is not the S-dual of
the worldsheet instanton correction $\lap G_g$. In particular,
they have a different functional form.
On the other hand, the sum over the $(p,q)$-strings in \eqref{eq:pq-sum}
is an exact statement and the $(p,q)$-string
contribution $\cG_N^{(\text{0-inst})}(M_{p,q})$ 
has exactly the same form as the contribution
of $(1,0)$-sting (or F-string) $\cG_N^{(\text{0-inst})}(M_{1,0})$.

\section{D3-instanton}\label{sec:D3-inst}
In this section, we study the large genus behavior of 
$G_g(\la)$ and $\til{G}_g(\til{\la})$.
It turns out that both of the genus expansions 
$\sum_g N^{2-2g}G_g(\la)$ and $\sum_g N^{1-2g}\til{G}_g(\til{\la})$
are Borel non-summable.
This means that the genus expansions receive non-perturbative corrections in $1/N$.
We argue that the non-perturbative correction to the resummation of $G_g(\la)$ corresponds to the magnetic D3-instanton, while
that to the resummation of $\til{G}_g(\til{\la})$ corresponds to the electric D3-instanton.
In fact, we observe that the Borel non-summability leads to the following discontinuities 
\begin{equation}
\begin{aligned}
 \text{Disc}\,\Biggl[ \sum_{g=0}^\infty N^{2-2g}G_g(\la) \Biggr]^{\text{Borel}}
&=\sum_{n=1}^\infty \frac{1}{n}G^{(\text{mag})}(N,nx), \\
\text{Disc}\,\Biggl[\sum_{g=0}^\infty N^{1-2g}\til{G}_g(\til{\la})\Biggr]^{\text{Borel}}
&=\sum_{n=1}^\infty G^{(\text{ele})}(N,ny),
\end{aligned}
\label{eq:genus-discontinuity} 
\end{equation}
where $G^{(\text{mag})}(N,x)$ and $G^{(\text{ele})}(N,y)$ with $x=\pi/\rt{\la}$ and $y=\pi/\rt{\til{\la}}$
denote the magnetic and electric D3-instanton corrections, respectively, whose explicit forms are given by
\begin{equation}
\begin{aligned}
 G^{(\text{mag})}(N,x)&=
\ri N\frac{(8Nx)^{\frac{3}{2}}}{\rt{2\pi}}(1+x^2)^{-\qu}
e^{-NA(x)}\sum_{k=0}^\infty 
\frac{g_k(x)}{\bigl[8Nx(1+x^2)^{\frac{3}{2}}\bigr]^k} ,\\
 G^{(\text{ele})}(N,y)&=
8\ri N^2
e^{-NA(y)}\sum_{k=0}^\infty 
\frac{h_k(y)}{\bigl[8Ny(1+y^2)^{\frac{3}{2}}\bigr]^k}.
\end{aligned}
\label{eq:G-mag}
\end{equation}
Here $A(z)=4\left[\text{arcsinh}\,(z)+z\rt{1+z^2}\right]$, and $g_k(z)$ and $h_k(z)$ are polynomials of $z$.

Curiously, the dual 't Hooft limit of the magnetic D3-instanton $G^{(\text{mag})}(N,x)$ formally reduces to the D-string instanton, which are computed in the previous section. Similarly, the electric D3-instanton reduces to the the worldsheet instanton in the 't Hooft limit.
That is to say, we observe that
\begin{equation}
\begin{aligned}
\sum_{n=1}^\infty \frac{1}{n}G^{(\text{mag})}(N,nx)
 &~\stackrel{\text{dual 't Hooft limit}}{ \xrightarrow{\hspace*{22mm}}}~ \sum_{g=0}^\infty N^{1-2g} \lap\til{G}_g(\til{\la}), \\
\sum_{n=1}^\infty G^{(\text{ele})}(N,ny)
 &~\stackrel{\text{'t Hooft limit}}{\xrightarrow{\hspace*{22mm}}}~ \sum_{g=0}^\infty N^{2-2g}\lap G_g(\la).
\end{aligned} 
\label{eq:resum-relation}
\end{equation}
Note that $G^{(\text{mag})}(N,x)$ is originally derived in the 't Hooft limit, not in the dual 't Hooft limit. One should understand it as an extrapolation from the 't Hooft limit to the dual 't Hooft limit via the very strong coupling limit, as shown in Figure~\ref{fig:N-gs}.
It is not obvious for us why the extrapolation to the dual 't Hooft limit of the magnetic D3-instanton reproduces the D-string instanton. 
However, as a result, the first line of \eqref{eq:resum-relation} agrees with the result in
\cite{Collier:2022emf}.

\subsection{Magnetic D3-instanton}
Let us first consider the resummation 
of $G_g(\la)$ in the large $\la$ regime.
In this regime, $G_g(\la)$ is expanded as
\begin{equation}
\begin{aligned}
 G_g(\la)\sim \sum_{m=0}^\infty c_{g,m}\la^{g-\hf-m}
\end{aligned} 
\label{eq:Gg-exp}
\end{equation}
where $c_{g,m}$ are some coefficients.
To see the large $g$ behavior in the large $\la$ regime, the terms with positive powers of $\la$ in \eqref{eq:Gg-exp} are relevant.
These terms come from the poles at right half-integer $s$ in \eqref{eq:GtilG-sint}.
Then we find
\begin{equation}
\begin{aligned}
 G_g^{+}=\sum_{j=1}^g \text{Res}_{s=j+\hf}
\biggl[-\frac{2\pi}{\sin\pi s}p(s)f_g(s)\Ga(1-s)\zeta(2-2s)\biggl(\frac{\la}{4}\biggr)^{s-1}\biggr],
\end{aligned} 
\end{equation}
where the superscript of $G_g^{+}$ refers to the positive power of $\la$.
The first few terms read
\begin{equation}
\begin{aligned}
 G_1^{+}&=-\frac{\sqrt{\lambda }}{16},\\
G_2^{+}&=\frac{\lambda ^{3/2}}{1536}-\frac{13 \sqrt{\lambda }}{8192},\\
G_3^{+}&=\frac{\lambda ^{5/2}}{98304}-\frac{25 \lambda ^{3/2}}{393216}+\frac{1533 \sqrt{\lambda
   }}{4194304},\\
G_4^{+}&=\frac{3 \lambda ^{7/2}}{5242880}-\frac{595 \lambda ^{5/2}}{100663296}+\frac{11473 \lambda
   ^{3/2}}{268435456}-\frac{1203917 \sqrt{\lambda }}{4294967296}.
\end{aligned} 
\end{equation}
We have computed $G_g^{+}$ up to $g=126$ 
\footnote{The data of $G_g^{+}$ are available 
upon request to the authors.}
and determined the large $g$ behavior of $G_g(\la)$ numerically
using the Richardson extrapolation
(see appendix \ref{app:richardson} for the details
of this method).
We find that the large $g$ asymptotic behavior of $G_g(\la)$
is given by
\begin{equation}
\begin{aligned}
 G_g(\la)\sim 
 16\left(\frac{x}{\pi}\right)^{\frac{3}{2}}(1+x^2)^{-\qu}
A(x)^{-2g-\hf}\sum_{k=0}^\infty g_k(x)\biggl[\frac{A(x)}{8x(1+x^2)^{\frac{3}{2}}}\biggr]^k \Ga\left(2g+\frac{1}{2}-k\right),
\end{aligned} 
\label{eq:Gg-asymp}
\end{equation}
where $x$ is defined by
\begin{equation}
\begin{aligned}
 x=\frac{\pi}{\rt{\la}}=\rt{\frac{\pi}{4g_sN}}.
\end{aligned} 
\label{eq:x-def}
\end{equation}
We find the first few terms of $g_k(x)$
\begin{equation}
\begin{aligned}
  g_0(x)&=1,\\
g_1(x)&=\frac{28 x^4}{3}+14 x^2+\frac{39}{8},\\
g_2(x)&=\frac{104 x^8}{9}+\frac{104 x^6}{3}+\frac{113 x^4}{2}+47 x^2+\frac{1785}{128},\\
g_3(x)&=\frac{1888 x^{12}}{405}+\frac{944 x^{10}}{45}+\frac{683 x^8}{5}+\frac{772
   x^6}{3}+\frac{7195 x^4}{32}+\frac{6951 x^2}{64}+\frac{22365}{1024},
\end{aligned} 
\end{equation}
and $A(x)$ in \eqref{eq:Gg-asymp} is given by
\begin{equation}
\begin{aligned}
 A(x)=4\left[\text{arcsinh}\,(x)+x\rt{1+x^2}\right].
\end{aligned} 
\label{eq:A-inst}
\end{equation}
As reviewed in appendix \ref{app:richardson}, the asymptotic behavior
\eqref{eq:Gg-asymp} leads to \eqref{eq:G-mag}, the discontinuity of the resummed genus expansion or twice of the non-perturbative correction in $1/N$.
Although the result \eqref{eq:G-mag} is a very complicated function of $x$, we are quite
confident of our numerics since we found more than 20 digits of agreement
in \eqref{eq:Gg-asymp} by applying the Richardson extrapolation to our data $\{G^{+}_g\}_{g\leq g_{\max}}$ with $g_{\max}=126$.

\begin{figure}[tb]
\centering
\includegraphics
[width=0.6\linewidth]{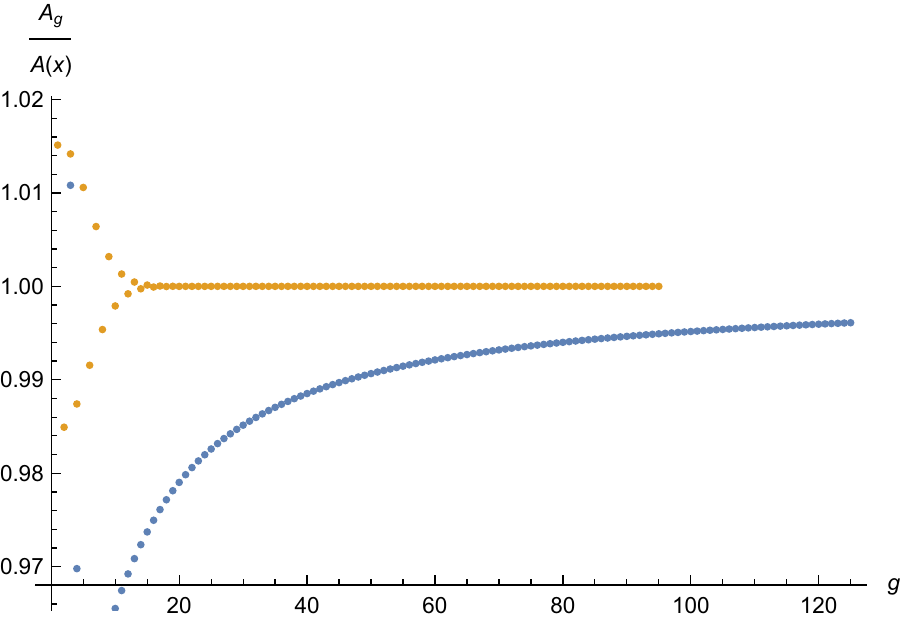}
\caption{
Plot of $A_g/A(x)$ for $\la=100$.
The blue dots represent $A_g/A(x)$
defined in \eqref{eq:Ag} while the 
orange dots represent its $30$-th Richardson transform
$A_g^{(30)}/A(x)$.
}
  \label{fig:ric}
\end{figure}

As a demonstration of the Richardson extrapolation, 
in Figure~\ref{fig:ric} we show the plot of the series
\begin{equation}
\begin{aligned}
 A_g=2g\rt{\frac{G_g^{+}}{G_{g+1}^{+}}},
\end{aligned} 
\label{eq:Ag}
\end{equation}
and its $30$-th Richardson transform $A^{(30)}_g$ for $\la=100$
(see \eqref{eq:Rich} for the definition of the
$k$-th Richardson transform).
From the discussion in appendix \ref{app:richardson}, this series approaches
$A(x)$ in the large $g$ limit.
Indeed, we find that the last term of this series $\{A^{(30)}_g\}
_{g\leq g_{\text{max}}-30}$
agrees with $A(x)$ in \eqref{eq:A-inst}
within $39$ digits accuracy
\begin{equation}
\begin{aligned}
 \frac{A^{(30)}_{g_{\max}-30}}{A(x)}-1=3.5904\times 10^{-40}.
\end{aligned} 
\end{equation}
We have also performed a similar analysis 
for $g_k(x)$ with various values of $\la$ and 
found a good numerical agreement using the method of Richardson extrapolation.

Note that the instanton action \eqref{eq:A-inst} is reminiscent of the action of 
the so-called ``giant Wilson loop''
in the $k$-th symmetric representation with large $k$ \cite{Drukker:2005kx,Gomis:2006sb,Gomis:2006im}.
In the bulk IIB string theory on $AdS_5\times S^5$, 
this giant Wilson loop corresponds to a D3-brane with $k$ units of electric flux
along its worldvolume. 
As shown in \cite{Drukker:2005kx}, 
the action of the giant Wilson loop is given by 
$\hf A(\ka)$ with the same function $A(x)$ in \eqref{eq:A-inst}, but the argument
$\ka$ is different from $x$ in \eqref{eq:x-def}
\begin{equation}
\begin{aligned}
 \ka=\frac{k\rt{\la}}{4N}=\rt{\frac{\pi k^2g_s}{4N}}.
\end{aligned} 
\label{eq:kappa}
\end{equation}
We can see that $\ka$ is related to $x$ by the formal replacement
$k^2g_s\to g_s^{-1}$.
This suggests that \eqref{eq:G-mag}
can be interpreted as the magnetic analogue of the D3-brane in 
\cite{Drukker:2005kx,Gomis:2006sb,Gomis:2006im}. Thus we call \eqref{eq:G-mag}
as the magnetic D3-instanton correction.

We observe that the constant term of $g_k(x)$
is equal to the coefficients appearing in $\lap\til{G}_0$ in \eqref{eq:tilGg}
\begin{equation}
\begin{aligned}
 g_k(0)=\til{a}_{0,k}.
\end{aligned} 
\end{equation}
This is not an accident. In fact, we find that
the D-string instanton 
correction $\lap \til{G}_g$ is reproduced 
from the magnetic D3-instanton in \eqref{eq:G-mag}
by taking the scaling limit
\begin{equation}
\begin{aligned}
 N\to\infty,\qquad x\to0,\qquad T_D=8Nx: \text{fixed}.
\end{aligned} 
\end{equation}
This is nothing but the dual 't Hooft limit.
More precisely, we further assume the strong coupling $T_D \gg 1$.
Therefore, we consider the regime where $\sqrt{\la} \sim N \to \infty$ and $\sqrt{\til{\la}} \gg 1$.
In this limit $G^{(\text{mag})}(N,x)$ in \eqref{eq:G-mag}
is expanded as
\begin{equation}
\begin{aligned}
&G^{(\text{mag})}\Bigl(N,\frac{T_D}{8N}\Bigr)
=\frac{\ri  T_D}{\rt{2\pi}}e^{-T_D} \sum_{g=0}^\infty N^{1-2g} \sum_{m=0}^\infty \til{a}_{g,m}T_D^{1/2+3g-m} \\
&=\frac{\ri  T_D}{\rt{2\pi}}e^{-T_D}
\Biggl[N\biggl( T_D^{1/2}+
\frac{39}{8}T_D^{-1/2}
+\frac{1785}{128}T_D^{-3/2}
+\frac{22365}{1024}T_D^{-5/2}+\cdots\biggr) \\
&\quad+N^{-1}\biggl(-\frac{1}{384}T_D^{7/2}
-\frac{17}{1024}T_D^{5/2}
+\frac{805}{16384}T_D^{3/2}+\cdots\biggr)\\
&\quad+N^{-3}\biggl(\frac{1}{294912}T_D^{13/2}
+\frac{43}{1310720}T_D^{11/2}
-\frac{2251}{20971520}T_D^{9/2}+\cdots\biggr)+\cO(N^{-5})
\Biggr].
\end{aligned} 
\end{equation}
This reproduces the single covering $(n=1)$ part of
the D-string instanton in \eqref{eq:lap-tilG}!
Conversely, if we use \eqref{eq:tilGg} we can go to the multi-covering structure:
\begin{equation}
\begin{aligned}
\sum_{g=0}^\infty N^{1-2g} \Delta \til{G}_g(\til{\la})
&=\sum_{g=0}^\infty N^{1-2g} \frac{\ri  T_D}{\rt{2\pi}}\sum_{n=1}^\infty e^{-nT_D} \sum_{m=0}^\infty \til{a}_{g,m}(nT_D)^{1/2+3g-m}\\
&=\sum_{n=1}^\infty \frac{1}{n} G^{(\text{mag})}(N,nx).
\end{aligned}
\label{eq:multi-mag}
\end{equation}
We conjecture that this gives the complete discontinuity of the genus resummation of $G_g(\la)$, and it implies the first line of \eqref{eq:genus-discontinuity}.

\subsection{Electric D3-instanton}
We can repeat the similar analysis for the second line of \eqref{eq:resum-relation}.
To see this, let us consider the large $g$ asymptotics of $\til{G}_g(\til{\la})$.
In the large $\til{\la}$ regime, $\til{G}_g(\til{\la})$ is expanded as
\begin{equation}
\begin{aligned}
 \til{G}_g(\til{\la})\sim \sum_{m=0}^\infty \til{c}_{g,m}\til{\la}^{g+\hf-m}.
\end{aligned} 
\label{eq:tilGg-exp}
\end{equation}
Note that the coefficients $\til{c}_{g,m}$ can be computed by the integral representation \eqref{eq:tilG_g}. Interestingly, these coefficients are simply related to the coefficients $c_{g,m}$ in $G_g(\la)$,
\begin{equation}
\begin{aligned}
\til{c}_{g,m}=\frac{c_{m,g}}{(4\pi)^{1+2g-2m}}.
\end{aligned}
\end{equation}
We found it by formally rewriting the genus expansion $\sum_{g} N^{2-2g}G_g(\la)$ at strong coupling in terms of $\til{\la}$.

As in the case of $G_g(\la)$, $\til{G}_g(\til{\la})$ is also approximated by
\begin{equation}
\begin{aligned}
 \til{G}_g^{+}=\sum_{j=1}^g \text{Res}_{s=j+\hf}
\biggl[-\frac{2\pi}{\sin\pi s}p(s)f_g(s)\Ga(s)\zeta(2s)\biggl(\frac{\til{\la}}{4\pi^2}\biggr)^{s}\biggr],
\end{aligned} 
\end{equation}
where the superscript of $\til{G}_g^{+}$ refers to the positive power of 
$\til{\la}$. The first few terms of $\til{G}_g^{+}$ read
\begin{equation}
\begin{aligned}
 \til{G}_1^{+}&=-\frac{3 \zeta (3)\til{\la}^{3/2}}{64 \pi ^3},\\
\til{G}_2^{+}&=\frac{45 \zeta (5) \til{\la}^{5/2}}{4096 \pi ^5}-\frac{39 \zeta (3) \til{\la}^{3/2}}{32768 \pi ^3},\\
\til{G}_3^{+}&=\frac{4725 \zeta (7) \til{\la}^{7/2}}{2097152 \pi ^7}-\frac{1125 \zeta (5) \til{\la}^{5/2}}{1048576 \pi
   ^5}+\frac{4599 \zeta (3) \til{\la}^{3/2}}{16777216 \pi ^3},\\
\til{G}_4^{+}&=
\frac{99225 \zeta (9) \til{\la}^{9/2}}{67108864 \pi ^9}-\frac{2811375 \zeta
   (7) \til{\la}^{7/2}}{2147483648 \pi ^7}+\frac{1548855 \zeta (5) \til{\la}^{5/2}}{2147483648 \pi
   ^5}-\frac{3611751 \zeta (3) \til{\la}^{3/2}}{17179869184 \pi ^3}.
\end{aligned} 
\end{equation}
We have computed $\til{G}_g^{+}$ up to $g=131$.\footnote{The data
of $\til{G}_g^{+}$ are available upon request to the authors.}
As in the previous subsection, we can 
determine the large $g$ behavior of $\til{G}_g^{+}$
numerically by the method of Richardson extrapolation
using our data of $\{\til{G}_g^{+}\}_{g\leq131}$.
The resulting non-perturbative correction is
given by \eqref{eq:G-mag},
where the action $A(y)$ is the same as \eqref{eq:A-inst}
but the argument $y$ is replaced by
\begin{equation}
\begin{aligned}
 y=\frac{\pi}{\rt{\til{\la}}}=\rt{\frac{\pi g_s}{4N}}.
\end{aligned} 
\end{equation}
One can see that $y$ is equal to $\ka$ in \eqref{eq:kappa} when $k=1$.
Thus it is natural to call $G^{\text{(ele)}}(N,y)$
as the electric D3-instanton correction.
The first few terms of the coefficients $h_k(y)$ in \eqref{eq:G-mag}
are given by\begin{equation}
\begin{aligned}
 h_0(y)&=1,\\
h_1(y)&=\frac{28 y^4}{3}+14 y^2+\frac{9}{2},\\
h_2(y)&=\frac{104 y^8}{9}+\frac{104 y^6}{3}+53 y^4+45 y^2+\frac{117}{8},\\
h_3(y)&=\frac{1888 y^{12}}{405}+\frac{944 y^{10}}{45}+\frac{1984 y^8}{15}+\frac{878 y^6}{3}+290
   y^4+\frac{585 y^2}{4}+\frac{489}{16}.
\end{aligned} 
\end{equation}
Again, we observe that the constant term of $h_k(y)$ 
is the same as the coefficients of $\lap G_0$ in \eqref{eq:lap-G0}
\begin{equation}
\begin{aligned}
 h_k(0)=a_k.
\end{aligned} 
\end{equation}
This suggests that $G^{(\text{ele})}(N,y)$ is related
to the worldsheet instanton studied in the previous section.
To see this, we take the scaling limit
\begin{equation}
\begin{aligned}
 N\to\infty,\qquad y\to0,\qquad T_F=8Ny: \text{fixed}.
\end{aligned} 
\end{equation}
Then $G^{(\text{ele})}(N,y)$ is expanded as
\begin{equation}
\begin{aligned}
 G^{(\text{ele})}\Bigl(N,\frac{T_F}{8N}\Bigr)&=\ri e^{-T_F}
\Biggl[8N^2 \left(1+
\frac{9}{2}T_F^{-1}
+\frac{117}{8}T_F^{-2}
+\frac{489}{16}T_F^{-3}+\cdots\right)\\
&+N^{0}\left(-\frac{1}{48}T_F^3
-\frac{3}{32}T_F^2
+\frac{77}{128}T_F+\cdots\right)\\
&+N^{-2}\left(\frac{1}{36864}T_F^6
+\frac{7}{40960}T_F^5
-\frac{857}{491520}T_F^4+\cdots\right)
+\cO(N^{-4})
\Biggr].
\end{aligned} 
\end{equation}
This reproduces the single covering ($n=1$) part of the worldsheet instanton 
corrections \eqref{eq:lap-G0}, \eqref{eq:lap-G1}, and \eqref{eq:lap-G23}.
The multi-covering structure is given by
\begin{equation}
\begin{aligned}
\sum_{g=0}^\infty N^{2-2g} \lap G_g(\la)&=\sum_{g=0}^\infty N^{2-2g}\cdot \ri \sum_{n=1}^\infty e^{-nT_F} \sum_{m=0}^\infty a_{g,m} (nT_F)^{3g-m} \\
&=\sum_{n=1}^\infty G^{(\text{ele})}(N,ny),
\end{aligned}
\end{equation}
which is almost parallel to \eqref{eq:multi-mag}. Again, 
we conjecture that this gives the complete discontinuity of the genus resummation of $\til{G}_g(\til{\la})$, 
and it implies the second line of \eqref{eq:genus-discontinuity}.

\subsection{Laplace-difference equation}
Following \cite{Marino:2008ya}, one can compute the magnetic and the
electric D3-instanton as a trans-series solution of the Laplace-difference equation
obeyed by the exact $\cG_N$ \cite{Dorigoni:2021guq}
\begin{equation}
\begin{aligned}
 (\lap_\tau-2)\cG_N=N^2(\cG_{N+1}+\cG_{N-1}-2\cG_{N})-N(\cG_{N+1}-\cG_{N-1}),
\end{aligned} 
\end{equation}
where $\lap_\tau$ is the Laplacian 
defined in \eqref{eq:lap-tau}. Here we show that the Laplace-difference equation fixes both the magnetic D3-instanton and the electric D3-instanton contributions, systematically.

Let us first consider the magnetic D3-instanton.
Since $G^{(\text{mag})}(N,x)$ is independent of $\tau_1$, 
when acting on $G^{(\text{mag})}(N,x)$ the Laplacian 
$\lap_\tau$ is replaced by
\begin{equation}
\begin{aligned}
\tau_2^2\frac{\del^2}{\del\tau_2^2}
G^{(\text{mag})}(N,x)
=\qu\left(x^2\del_x^2-x\del_x\right)G^{(\text{mag})}(N,x),
\end{aligned} 
\end{equation}
where $x$ is related to $\tau_2$ by
\begin{equation}
\begin{aligned}
 x=\rt{\frac{\pi \tau_2}{4N}}.
\end{aligned} 
\end{equation}
Thus the magnetic D3-instanton function
should satisfy
\begin{equation}
\begin{aligned}
 &\left(\frac{x^2\del_x^2-x\del_x}{4}-2\right)G^{(\text{mag})}(N,x)\\
=&N^2\Bigl[G^{(\text{mag})}(N+1,x_+)+G^{(\text{mag})}(N-1,x_{-})
-2G^{(\text{mag})}(N,x)\Bigr]\\
&-N\Bigl[G^{(\text{mag})}(N+1,x_+)-G^{(\text{mag})}(N-1,x_{-})\Bigr],
\end{aligned} 
\label{eq:diffeq-mag}
\end{equation}
where
\begin{equation}
\begin{aligned}
 x_\pm=x(1\pm N^{-1})^{-\hf}.
\end{aligned} 
\end{equation}
At the leading order in the large $N$ expansion
of \eqref{eq:diffeq-mag}, the action $A(x)$
should satisfy
\begin{equation}
\begin{aligned}
 \hf x\del_xA(x)=2\sinh\left(\hf A(x)-\qu x\del_xA(x)\right).
\end{aligned} 
\label{eq:A-eq}
\end{equation}
One can check that $A(x)$ in \eqref{eq:A-inst} indeed satisfies
the above equation \eqref{eq:A-eq}.
One can also check that $G^{(\text{mag})}(N,x)$ in \eqref{eq:G-mag}
satisfies \eqref{eq:A-eq} order by order in the $1/N$ expansion.
Alternatively, we can compute $g_k(x)$ in \eqref{eq:G-mag}
recursively from the Laplace-difference equation \eqref{eq:A-eq}.
However, the constant term of $g_k(x)$ is not fixed by the 
equation \eqref{eq:A-eq} alone;
we have to supply the information of $g_k(0)=\til{a}_{0,k}$ as an input. 
With this input, we can compute $g_k(x)$ recursively up to any 
desired order.
Using this result of $g_k(x)$, we observed that
the highest and the next highest degree terms of $g_k(x)$
have the structure
\begin{equation}
\begin{aligned}
 g_k(x)=c_k\left(x^{4k}+\frac{3k}{2}x^{4k-2}\right)+\cO(x^{4k-4}),
\end{aligned} 
\end{equation}
where $c_k$ is determined by the generating function
\begin{equation}
\begin{aligned}
 \sum_{k=0}^\infty \frac{c_k}{(8N)^k}=\frac{1+N}{N}
\left(\frac{\Ga(N+1)}{\rt{2\pi N}N^N e^{-N}}\right)^2.
\end{aligned} 
\end{equation}  

Similarly, the electric D3-instanton can be
characterized as a trans-series solution of the following Laplace-difference equation
\begin{equation}
\begin{aligned}
  &\left(\frac{y^2\del_y^2+3y\del_y}{4}-2\right)G^{(\text{ele})}(N,y)\\
=&N^2\Bigl[G^{(\text{ele})}(N+1,y_+)+G^{(\text{ele})}(N-1,y_{-})
-2G^{(\text{ele})}(N,y)\Bigr]\\
&-N\Bigl[G^{(\text{ele})}(N+1,y_+)-G^{(\text{ele})}(N-1,y_{-})\Bigr],
\end{aligned} 
\label{eq:diffeq-ele}
\end{equation}
where $y_\pm=y(1\pm N^{-1})^{-\hf}$. Again, one can check 
that $G^{(\text{ele})}(N,y)$ in \eqref{eq:G-mag}
satisfies \eqref{eq:diffeq-ele} order by order in the $1/N$ expansion.
Alternatively, one can compute the coefficient $h_k(y)$
recursively from the Laplace-difference equation \eqref{eq:diffeq-ele}.
However, as in the case of $g_k(x)$, the constant term $h_k(0)$
is not determined by the equation \eqref{eq:diffeq-ele} alone, and we should supply
the information of $h_k(0)=a_k$ as input.
With this input, we can compute $h_k(y)$ up to any desired order from the
equation \eqref{eq:diffeq-ele}.

\section{Conclusions and outlook}\label{sec:conclusion}
In this paper, we studied the large $N$ limit of the integrated correlator $\cG_N$. We rewrote the result in \cite{Dorigoni:2021bvj,Dorigoni:2021guq} as a sum over $(p,q)$-string contributions. Due to the $SL(2,\mathbb{Z})$ duality, it is natural that all the $(p,q)$-strings contribute democratically.
We found the recursion relation for $f_g(s)$ to compute the large $N$ expansions of $\cG_N^{\text{(0-inst)}}(M_{1,0})$ up to very high genera.

We showed that the genus expansions $\sum_g N^{2-2g}G_g(\la)$ and $\sum_g N^{1-2g}\til{G}_g(\til{\la})$ receive non-perturbative corrections,
which correspond to the magnetic D3-instanton correction $G^{(\text{mag})}(N,x)$ 
and
the electric D3-instanton correction $G^{(\text{ele})}(N,y)$, respectively. 
The forms of these non-perturbative corrections are constrained by the large order behaviors of $G_g(\la)$ and $\til{G}_g(\til{\la})$, and given by \eqref{eq:G-mag}. Interestingly, $G^{(\text{mag})}(N,x)$ reproduces the D-string instanton correction in the dual 't Hooft limit
while $G^{(\text{ele})}(N,y)$ reproduces the worldsheet instanton correction in the 't Hooft limit. We confirmed that the same corrections are fixed by the Laplace-difference equation for the integrated correlator. 

There are many open questions.
In this paper, we have not studied the D-instanton corrections 
in detail
(see appendix~\ref{app:1inst} for the exact 
result of one-instanton contribution at finite $N$). 
As we emphasized at the end of section~\ref{sec:pq},
D-instanton corrections cannot be seen in 
the $(p,q)$-string contribution $\cG_N^{(\text{0-inst})}(M_{p,q})$
with fixed $(p,q)$.
We have seen that the $(1,0)$-string contribution $\cG_N^{(\text{0-inst})}(M_{1,0})$
is naturally organized into $G(\la)$ and $\til{G}(\til{\la})$, whose
non-perturbative corrections $e^{-2\rt{\la}}$ and $e^{-2\rt{\til{\la}}}$
are interpreted as the worldsheet instanton and the D-string instanton, 
respectively.
However, we do not see the mixed contribution of these 
instantons.
On the other hand, in the grand partition function of
ABJM theory on $S^3$, 
it is known that there are ``bound states'' of worldsheet instantons and 
membrane instantons \cite{Hatsuda:2013gj}. 
It turns out that these bound states can be absorbed into the worldsheet instantons
by a certain redefinition of the chemical potential,
which corresponds to the quantum mirror map of the
topological string on local 
$\mathbb{P}^1\times \mathbb{P}^1$ \cite{Hatsuda:2013oxa}.
It would be interesting to see if such a structure exists in the 
integrated correlator of $\cN=4$ SYM as well.

In a recent paper \cite{Dorigoni:2022zcr}, 
the lattice sum representation of the integrated correlator \eqref{eq:def-GN}
is generalized to the $\cN=4$ SYM with gauge groups $SO(N)$ and $Sp(N)$.
It would be interesting to repeat our analysis with these gauge groups
and see
how the unoriented worldsheet contributions fit into the whole picture.
It would also be interesting to 
generalize our analysis to another type of integrated correlator 
$\del_m^4\log Z\big|_{m=0}$.
As discussed in \cite{Chester:2020vyz,Collier:2022emf},
in this correlator there appear
certain generalized Eisenstein series, which satisfy inhomogeneous Laplace
equations on the upper-half $\tau$-plane.
We leave these generalization of our analysis for future works.

\acknowledgments
We are grateful to Kazuhiro Sakai for collaboration at the initial stage of this
work.
The work of YH is supported
in part by JSPS KAKENHI Grant No. 18K03657 and 22K03641.
The work of KO is supported
in part by JSPS Grant-in-Aid for Transformative Research Areas (A) 
``Extreme Universe'' No. 21H05187 and JSPS KAKENHI Grant No. 22K03594.

\appendix
\section{Exact results at finite $N$}\label{app:exact}

\subsection{Relation between $B_N(t)$ and $K_N(x)$}

The zero instanton part $\cG_N^{(\text{0-inst})}$
is written as
\begin{equation}
\begin{aligned}
 \cG_N^{(\text{0-inst})}
&=\int_0^\infty dt B_N(t)\sum_{n\in\mZ}\frac{1}{\rt{g_st}}e^{-\frac{\pi n^2}{g_st}}.
\end{aligned} 
\label{eq:0inst-B}
\end{equation} 
On the other hand, $\cG_N^{(\text{0-inst})}$ is also written as
\cite{Russo:2013kea}
\begin{equation}
\begin{aligned}
 \cG_N^{(\text{0-inst})}
=\int_0^\infty dw \frac{w K_N(x)}{\sinh^2 w},\quad 
x=\frac{g_sw^2}{\pi}.
\end{aligned} 
\label{eq:0inst-K}
\end{equation}
Here $K_N(x)$ is given by a certain correlator
in the Gaussian matrix model
\begin{equation}
\begin{aligned}
 K_N(x)=-\hf \del_x x^2\del_x \text{SFF}(x),
\end{aligned} 
\end{equation}
where
\begin{equation}
\begin{aligned}
 \text{SFF}(x)=\left\bra\Tr e^{\ri wM}\Tr e^{-\ri wM}\right\ket
=\frac{\int dM e^{-\frac{2\pi}{g_s}\Tr M^2}\Tr e^{\ri wM}\Tr e^{-\ri wM}}{
\int dM e^{-\frac{2\pi}{g_s}\Tr M^2}}.
\end{aligned} 
\label{eq:SFF-def}
\end{equation}
$\text{SFF}(x)$ in \eqref{eq:SFF-def} is known as the spectral form factor
if we regard $w$ as ``time'' and $M$ as a random Hamiltonian. 
The matrix model correlator \eqref{eq:SFF-def}
is naturally decomposed into the disconnected part
and the connected part
\begin{equation}
\begin{aligned}
 \text{SFF}_{\text{dis}}(x)&=\left\bra\Tr e^{\ri wM}\right\ket
\left\bra\Tr e^{-\ri wM}\right\ket,\\
\text{SFF}_{\text{conn}}(x)&=\left\bra\Tr e^{\ri wM}\Tr e^{-\ri wM}\right\ket_{\text{conn}}.
\end{aligned} 
\end{equation}
It is known that $\text{SFF}_{\text{dis}}(x)$
and $\del_x\text{SFF}_{\text{conn}}(x)$ can be written in a closed form
in terms of the Laguerre polynomials
\cite{Brezin1997,Okuyama:2018gfr,Okuyama:2018yep}
\begin{equation}
\begin{aligned}
 \text{SFF}_{\text{dis}}(x)&=e^{-x}L_{N-1}^{(1)}(x)^2,\\
\del_x \text{SFF}_{\text{conn}}(x)&=Ne^{-x}
\Bigl[L_{N-1}^{(1)}(x)L_{N-1}(x)-L_{N-2}^{(1)}(x)L_{N}(x)\Bigr].
\end{aligned} 
\label{eq:SFF-Lag}
\end{equation}
It turns out that $\text{SFF}_{\text{dis}}(x)$ and $\text{SFF}_{\text{conn}}(x)$
are not independent; they are related by \cite{Brezin1997}
\begin{equation}
\begin{aligned}
 \del_x^2\text{SFF}_{\text{conn}}(x)=-\text{SFF}_{\text{dis}}(x).
\end{aligned} 
\label{eq:rel-SFF}
\end{equation}

As discussed in \cite{Dorigoni:2022zcr}, $B_N(t)$ and $K_N(x)$ are related
by the Laplace transformation
\begin{equation}
\begin{aligned}
 B_N(t)=-\left(t\del_t+\hf\right)\int_0^\infty dx e^{-tx}K_N(x)
=\int_0^\infty dx e^{-tx}\left(tx-\hf\right)K_N(x).
\end{aligned} 
\label{eq:BtoK}
\end{equation}
Indeed, we can easily check that
plugging \eqref{eq:BtoK} into \eqref{eq:0inst-B}
we recover \eqref{eq:0inst-K}
\begin{equation}
\begin{aligned}
 \cG_N^{(\text{0-inst})}&=\int_0^\infty \frac{2g_swdw}{\pi}K_N(x)\int_0^\infty dt
\left(t\frac{g_sw^2}{\pi}-\hf\right)e^{-t\frac{g_sw^2}{\pi}} \sum_{n\in\mZ}\frac{1}{\rt{g_st}}e^{-\frac{\pi n^2}{g_st}}\\
&=\int_0^\infty dw w K_N(x)\sum_{n\in\mZ} 2|n|e^{-2|n|w}\\
&=\int_0^\infty dw\frac{wK_N(x)}{\sinh^2w}.
\end{aligned} 
\end{equation}
Using the relation between $B_N(t)$ and $K_N(x)$ in
\eqref{eq:BtoK}, one can easily show the basic properties of $B_N(t)$
found in \cite{Dorigoni:2021guq}.
The first property is
\begin{equation}
\begin{aligned}
 \int_0^\infty\frac{dt}{\rt{t}}B_N(t)=0,
\end{aligned} 
\end{equation}
which follows from 
\begin{equation}
\begin{aligned}
 \int_0^\infty\frac{dt}{\rt{t}} e^{-tx}\left(tx-\hf\right)=
-\int_0^\infty dt\del_t(\rt{t}e^{-tx})
=0.
\end{aligned} 
\end{equation}
The second property is
\begin{equation}
\begin{aligned}
 \int_0^\infty dtB_N(t)=\qu N(N-1).
\end{aligned} 
\label{eq:BN-int}
\end{equation}
This can be shown as
\begin{equation}
\begin{aligned}
 &\int_0^\infty dt \int_0^\infty dx e^{-tx}\left(tx-\hf\right)K_N(x)\\
=&\int_0^\infty dx\frac{1}{2x}K_N(x)\\
=&-\qu \int_0^\infty dx\frac{1}{x}\del_xx^2\del_x\text{SFF}(x)\\
=&\frac{\text{SFF}(0)-\text{SFF}(\infty)}{4},
\end{aligned} 
\end{equation}
where we performed the integration by parts in the last
equality.
It is well known that the early time and the late time value of 
the spectral form factor are given by
\begin{equation}
\begin{aligned}
 \text{SFF}(0)=N^2,\quad \text{SFF}(\infty)=N,
\end{aligned} 
\end{equation}
which reproduces \eqref{eq:BN-int}.

Using the explicit form of $\text{SFF}(x)$
in \eqref{eq:SFF-Lag},
one can prove the relation \eqref{eq:BtoK}
by directly computing the Laplace transform of $K_N(x)$.\footnote{
This part is based on a discussion with Kazuhiro Sakai.
We would like to thank him for sharing his unpublished note.}
To this end, it is natural to define 
\begin{equation}
\begin{aligned}
 B_{\text{dis}}(t)&=
\int_0^\infty dx e^{-tx}\left(tx-\hf\right)\left(-\hf\del_xx^2\del_x\text{SFF}_{\text{dis}}(x)\right),\\
 B_{\text{conn}}(t)&=\int_0^\infty dx e^{-tx}\left(tx-\hf\right)\left(-\hf\del_xx^2\del_x\text{SFF}_{\text{conn}}(x)\right).
\end{aligned} 
\end{equation}
Let us first consider $B_{\text{dis}}(t)$
\begin{equation}
\begin{aligned}
B_{\text{dis}}(t)
&=\hf(t\del_t+1/2)\int_0^\infty dxe^{-tx}\del_xx^2\del_xe^{-x}L_{N-1}^{(1)}(x)^2\\
&=\hf (t\del_t+1/2)\int_0^\infty dxe^{-(1+t)x}(-2tx+t^2x^2)L_{N-1}^{(1)}(x)^2\\
&=-\hf (t\del_t+1/2)(2t+t^2\del_t)I
\end{aligned} 
\end{equation}
where $I$ is given by
\begin{equation}
\begin{aligned}
 I&=\int_0^\infty dxe^{-(1+t)x} xL_{N-1}^{(1)}(x)^2.
\end{aligned} 
\end{equation}
Using the formula in \cite{poh2001some}, this is evaluated as
\begin{equation}
\begin{aligned}
 I=N^2\frac{t^{2N-2}}{(1+t)^{2N}}{}_2F_1(1-N,1-N,2;t^{-2}).
\end{aligned} 
\end{equation}
$B_{\text{conn}}(t)$ can be computed in a similar manner.
After some algebra, we find
 \begin{equation}
\begin{aligned}
 B_{\text{dis}}(t)&=\frac{3Nt^{2N-2}(1-t^2-2Nt+2N^2t^2)}{4(t-1)(1+t)^{2N+1}}X\\
&+\frac{Nt^{2N-2}\bigl[-3(1-t^2)^2+12Nt-24N^2t^2-4Nt^3+16N^3t^3\bigr]}{4(t-1)^2(1+t)^{2N+2}}Y,\\
B_{\text{conn}}(t)&=\frac{3Nt^{2N-2}(1-t^2+2Nt-2N^2)}{4(t-1)(1+t)^{2N+1}}X\\
&-\frac{Nt^{2N-2}\bigl[-3(1-t^2)^2+12Nt^3-24N^2t^2-4Nt+16N^3t\bigr]}{4(t-1)^2(1+t)^{2N+2}}Y,
\end{aligned} 
\label{eq:Bdis-conn}
\end{equation}
where
\begin{equation}
\begin{aligned}
 X&=t^{3-2N}{}_2F_1(-N,2-N,2;t^2)+{}_2F_1(-N,2-N,2;t^{-2}),\\
Y&=t^{3-2N}{}_2F_1(-N,2-N,2;t^2)-{}_2F_1(-N,2-N,2;t^{-2}).
\end{aligned} 
\label{eq:XY-def}
\end{equation}
We observe that $B_{\text{dis}}(t)$ and $B_{\text{conn}}(t)$ are related
by
\begin{equation}
\begin{aligned}
 B_{\text{conn}}(t)=t^{-1}B_{\text{dis}}(t^{-1}).
\end{aligned} 
\end{equation}
Adding $B_{\text{dis}}(t)$ and $B_{\text{conn}}(t)$ in \eqref{eq:Bdis-conn}, 
we find a simple expression of $B_N(t)$
\begin{equation}
\begin{aligned}
 B_N(t)&=B_{\text{dis}}(t)+B_{\text{conn}}(t)
=\frac{N(N^2-1)t^{2N-2}}{(1+t)^{2N+1}}\left[\frac{3}{2}(1+t)X
+\frac{4Nt}{t-1}Y\right].
\end{aligned} 
\label{eq:BfromK}
\end{equation}
One can show that
$X$ and $Y$ in \eqref{eq:XY-def} are related to the Jacobi polynomials
as
\begin{equation}
\begin{aligned}
 (N+1)t^{2N-2}X&=(1-t^2)^N\left[(t-1)P_N^{(1,-2)}\left(\frac{1+t^2}{1-t^2}\right)
+P_N^{(1,-1)}\left(\frac{1+t^2}{1-t^2}\right)\right],\\
(N+1)t^{2N-2}Y&=(1-t^2)^N\left[(t+1)P_N^{(1,-2)}\left(\frac{1+t^2}{1-t^2}\right)
-P_N^{(1,-1)}\left(\frac{1+t^2}{1-t^2}\right)\right],
\end{aligned} 
\end{equation}
which implies the equivalence of \eqref{eq:BfromK} and \eqref{eq:BN}.

\subsection{Closed form of $c(s,N)$}
The spectral function $c(s,N)$ is defined as the
expansion coefficient of $B_N(t)$ in \eqref{eq:B-cs}.
From \eqref{eq:BtoK}, one can show that $c(s,N)$
is given by the Mellin transform of $K_N(x)$
\begin{equation}
\begin{aligned}
 c(s,N)=\Bigl(s-\hf\Bigr)\int_0^\infty dx \frac{x^{s-1}}{\Ga(s)}K_N(x).
\end{aligned} 
\end{equation}
From \eqref{eq:SFF-Lag} and \eqref{eq:rel-SFF}, $K_N(x)$
is written as
\begin{equation}
\begin{aligned}
  K_N(x)=-\hf \del_x \left(x^2\del_x\Bigl[e^{-x}L_{N-1}^{(1)}(x)^2\Bigr] \right)
+\hf \del_x \left(x^2\int_{\infty}^x dy e^{-y}L_{N-1}^{(1)}(y)^2\right).
\end{aligned} 
\label{eq:KN-exp}
\end{equation}
Then, by the partial integration,
the Mellin transform of \eqref{eq:KN-exp} becomes
\begin{equation}
\begin{aligned}
 \int_0^\infty dx\,x^{s-1}K_N(x)=\frac{s-1}{2}\int_0^\infty dx\left(-sx^{s-1}
+\frac{x^{s+1}}{s+1}\right)e^{-x}L_{N-1}^{(1)}(x)^2.
\end{aligned} 
\end{equation}
Using the relation
\begin{equation}
\begin{aligned}
 &\int_0^\infty dx x^\al e^{-x}L_n^{(\al)}(x)L_m^{(\al)}(x)=
\frac{\Ga(n+1+\al)}{\Ga(n+1)}\cob_{n,m},\\
&L_n^{(\al)}(x)=\sum_{i=0}^n \binom{\al-\bt+n-i-1}{n-i}L_i^{(\bt)}(x),
\end{aligned} 
\end{equation}
we find
\begin{equation}
\begin{aligned}
 &\int_0^\infty dx\,x^{s-1}K_N(x)\\
=&\frac{s-1}{2}\sum_{i=0}^{N-1}\left[
-s\binom{N-s-i}{N-1-i}^2\frac{\Ga(i+s)}{\Ga(i+1)}+\frac{1}{s+1}
\binom{N-s-i-2}{N-1-i}^2\frac{\Ga(i+s+2)}{\Ga(i+1)}\right].
\end{aligned}
\end{equation}
This summation can be written in terms of the hypergeometric function
${}_3F_2$. Finally, we find
the exact form of $c(s,N)$
at finite $N$
\begin{equation}
\begin{aligned}
 c(s,N)&= 
\frac{s(s-1/2)}{2\Ga(N)^2\Ga(2-s)\Ga(1-s)}
\Biggl[\Ga(N+1-s)^2\,{}_3F_2\bigl(\{1-N,1-N,s\},\{s-N,s-N\},1\bigr)\\
&\quad
-(s-1)^2s^2\Ga(N-1-s)^2\,{}_3F_2\bigl(\{1-N,1-N,s+2\},\{s+2-N,s+2-N\},1\bigr)\Biggr].
\end{aligned}
\label{eq:c-finite-N} 
\end{equation}
At finite $N$, $c(s,N)$ is a polynomial in $s$. For instance,
\begin{equation}
\begin{aligned}
 c(s,2)&=\hf(1-2s)^2s(s-1),\\
c(s,3)&=\qu (1-2s)^2s(s-1)(6-s+s^2),\\
c(s,4)&=\frac{1}{24}(1-2s)^2s(s-1)(4-s+s^2)(18-s+s^2).
\end{aligned}
\label{eq:c-finite-N-2} 
\end{equation}
Note also that $c(s,N)$ has a symmetry
\begin{equation}
\begin{aligned}
 c(1-s,N)=c(s,N).
\end{aligned} 
\end{equation}

\subsection{Exact one-instanton contribution at finite $N$}\label{app:1inst}
From \eqref{eq:k-inst}, the one-instanton contribution is written as
\begin{equation}
\begin{aligned}
 \cG^{(\text{1-inst})}_N=\int_0^\infty\frac{dt}{\rt{g_st}}B_N(t)e^{-\frac{\pi}{g_s}
(t+t^{-1})}.
\end{aligned} 
\label{eq:1-inst}
\end{equation}
In order to evaluate this integral, it is convenient to 
perform the change of integration variable
\begin{equation}
\begin{aligned}
 t=e^{2x}.
\end{aligned} 
\end{equation}
Then \eqref{eq:1-inst} becomes
\begin{equation}
\begin{aligned}
 \cG^{(\text{1-inst})}_N=\frac{2}{\rt{g_s}}\int_{-\infty}^\infty
dx \,e^{x}B_N(e^{2x})e^{-\frac{2\pi}{g_s}\cosh 2x}.
\end{aligned} 
\end{equation}
Using \eqref{eq:BfromK}, one can show that
$e^{x}B_N(e^{2x})$ is expanded as
\begin{equation}
\begin{aligned}
e^{x}B_N(e^{2x})= &\frac{N(N^2-1)}{\rt{\pi}(\cosh x)^{2N+1}}\sum_{j=0}^{N-1}
(\cosh x)^{2j}(-1)^{N+j+1}
\Ga(N+1/2-j)\Ga(N-1)\\
\times &\Biggl[\frac{9}{4\Ga(j)\Ga(N-j)\Ga(N+3-j)}+
\frac{N}{\Ga(j+1)\Ga(N-1-j)\Ga(N+2-j)}
\Biggr].
\end{aligned} 
\end{equation}
Plugging this into \eqref{eq:1-inst}, we find
\begin{equation}
\begin{aligned}
 \cG^{(\text{1-inst})}_N=&\frac{2\Ga(N+2)}{\rt{g_s}}
 e^{-\frac{2\pi}{g_s}}\sum_{j=0}^{N-1}
(-1)^{N+j+1}U\Bigl(\hf,\hf+j-N,\frac{4\pi}{g_s}\Bigr)
\Ga\Bigl(N+\hf-j\Bigr)\\
\times & \Biggl[\frac{9}{4\Ga(j)\Ga(N-j)\Ga(N+3-j)}+
\frac{N}{\Ga(j+1)\Ga(N-1-j)\Ga(N+2-j)}
\Biggr],
\end{aligned} 
\label{eq:1-inst-exact}
\end{equation}
where $U(a,b,z)$ denotes the Toriconi's confluent hypergeometric function.
For instance, when $N=2$ and $N=3$ 
\eqref{eq:1-inst-exact} becomes
\begin{equation}
\begin{aligned}
 \cG_2^{(\text{1-inst})}
&=\frac{12\pi^2}{g_s^2}e^{-\frac{2\pi}{g_s}}-\frac{3\pi^2}{g_s^{3/2}}
\left(1+\frac{8\pi}{g_s}\right)
e^{\frac{2\pi}{g_s}}\text{Erfc}\left(2\rt{\frac{\pi}{g_s}}\right),\\
\cG_3^{(\text{1-inst})}
&=
\frac{3\pi^2}{g_s^2}\left(37+\frac{40\pi}{g_s}\right)e^{-\frac{2\pi}{g_s}}
-\frac{3\pi^2}{8g_s^{3/2}}\left(27+\frac{336\pi}{g_s}+\frac{320\pi^2}{g_s^2}\right)
e^{\frac{2\pi}{g_s}}\text{Erfc}\left(2\rt{\frac{\pi}{g_s}}\right),
\end{aligned} 
\label{eq:1-inst-N23}
\end{equation}
where $\text{Erfc}(z)$ denotes the complementary error function.
This agrees with the result in \cite{Dorigoni:2021guq}.

From the exact result
\eqref{eq:1-inst-exact}, we find the small $g_s$ expansion 
of $\cG^{(\text{1-inst})}_N$
\begin{equation}
\begin{aligned}
  \cG^{(\text{1-inst})}_N=&-\frac{3\Ga(N-1/2)}{4\rt{\pi}\Ga(N-1)}
e^{-\frac{2\pi}{g_s}}\Biggl[1+\frac{3(N-4)}{8(2N-3)}\frac{g_s}{\pi}-
\frac{15N(N+1)}{128(2N-3)(2N-5)}\frac{g_s^2}{\pi^2}\\
+&\frac{105N(N+1)(N-6)}{1024(2N-3)(2N-5)(2N-7)}\frac{g_s^3}{\pi^3}
-\frac{4725N(N+1)(N^2-11N+38)}{32768(2N-3)(2N-5)(2N-7)(2N-9)}\frac{g_s^4}{\pi^4}
+\cdots\Biggr].
\end{aligned} 
\end{equation}
\section{Relation between $B_{\text{dis}},B_{\text{conn}}$ and $G_0(\la),\til{G}_0(\til{\la})$}\label{app:dis-conn}
\begin{figure}[htb]
\centering
\subcaptionbox{$G_0(\la)$ \label{sfig:conn-g0}}{\includegraphics
[width=0.4\linewidth]{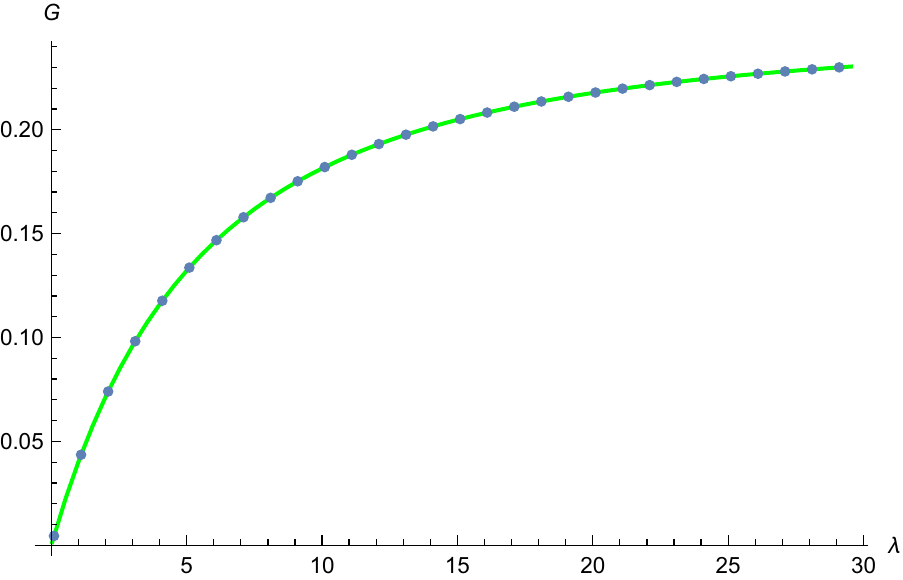}}
\hskip8mm
\subcaptionbox{$\til{G}_0(\til{\la})$ \label{sfig:dis-g0}}{\includegraphics
[width=0.4\linewidth]{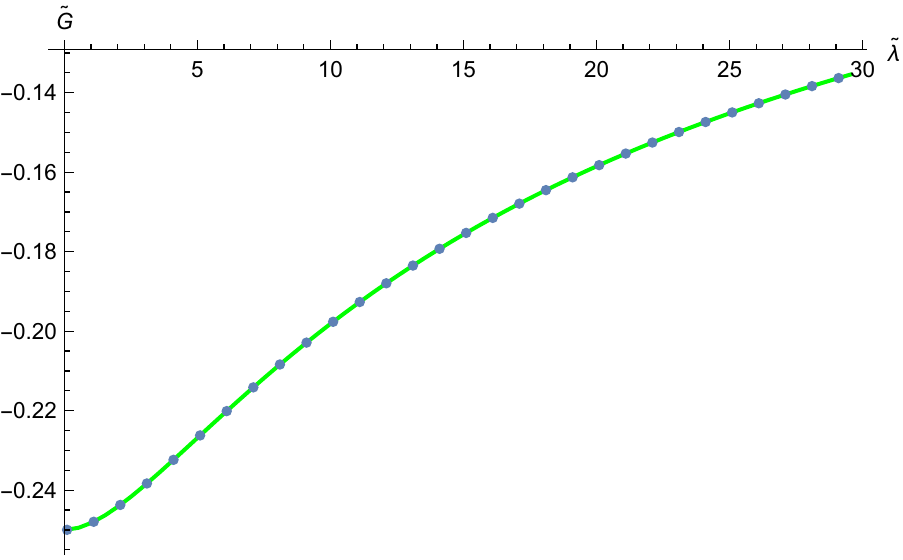}}
  \caption{
Plot 
of 
\subref{sfig:conn-g0} $G_0(\la)$ and 
\subref{sfig:dis-g0} $\til{G}_0(\til{\la})$.
The solid green curves represent the left hand side of
\eqref{eq:G0-approx} for $N=20$, while the blue dots represent
the genus-zero result
in \eqref{eq:G0-int}.
}
  \label{fig:G-genus0}
\end{figure}

From the matrix model picture, there is a natural decomposition
of the zero-instanton part into the disconnected and the connected part
\begin{equation}
\begin{aligned}
 \cG_{\text{dis}}^{(\text{0-inst})}&=\int_0^\infty dt
B_{\text{dis}}(t)\vartheta_3(e^{-2\pi g_st}),\\
\cG_{\text{conn}}^{(\text{0-inst})}&=\int_0^\infty dt
B_{\text{conn}}(t)\vartheta_3(e^{-2\pi g_st}),
\end{aligned} 
\end{equation}
where $B_{\text{dis}}(t)$ and $B_{\text{conn}}(t)$ are given by
\eqref{eq:Bdis-conn}.
We observe that
$\cG_{\text{dis}}^{(\text{0-inst})}$ 
approaches $N^2G_0(\la)$
in the 't Hooft limit,
and $\cG_{\text{conn}}^{(\text{0-inst})}$
approaches $N\til{G}_0(\til{\la})$
in the dual 't Hooft limit
\begin{equation}
\begin{aligned}
 \lim_{N\to\infty}N^{-2}\cG_{\text{dis}}^{(\text{0-inst})}\Big|_{g_s=\frac{\la}{4\pi N}}&= G_0(\la),\\
\lim_{N\to\infty}N^{-1}\cG_{\text{conn}}^{(\text{0-inst})}\Big|_{g_s=\frac{4\pi N}{\til{\la}}}
&= \til{G}_0(\til{\la}),
\end{aligned} 
\label{eq:G0-approx}
\end{equation}
where $G_0(\la)$ and $\til{G}_0(\til{\la})$ are given by
\cite{Russo:2013kea,Binder:2019jwn,Collier:2022emf}
\begin{equation}
\begin{aligned}
 G_0(\la)&=\int_0^\infty dw w
\frac{J_1(w\rt{\la}/\pi)^2-J_2(w\rt{\la}/\pi)^2}{\sinh^2w},\\
\til{G}_0(\til{\la})&=\frac{\til{\la}}{2\pi}\int_0^1 dw\frac{1}
{\sinh^2(w\rt{\til{\la}})}\frac{w^2(4w^2-3)}{\rt{1-w^2}}.
\end{aligned} 
\label{eq:G0-int}
\end{equation}
As we can see from Figure~\ref{fig:G-genus0}, 
\eqref{eq:G0-approx} is a good approximation at large $N$.

The relation between $\cG_{\text{dis}}^{(\text{0-inst})}$ and 
$G_0(\la)$ can be understood as follows.
In the large $N$ limit, $\bra\Tr e^{\pm\ri wM}\ket$ becomes
\begin{equation}
\begin{aligned}
 \bra\Tr e^{\pm\ri wM}\ket=N\frac{2J_1(\rt{4Nx})}{\rt{4Nx}},
\end{aligned} 
\end{equation}
and the disconnected part of $K_N(x)$ is given by
\begin{equation}
\begin{aligned}
 -\hf\del_xx^2\del_x\left(N\frac{2J_1(\rt{4Nx})}{\rt{4Nx}}\right)^2
=N^2\Bigl[J_1(\rt{4Nx})^2-J_1(\rt{4Nx})^2\Bigr],
\end{aligned} 
\end{equation}
which reproduces $G_0(\la)$ in \eqref{eq:G0-int}.

On the other hand, 
we do not have a clear understanding of the reason why 
the connected part $\cG_{\text{conn}}^{(\text{0-inst})}$
corresponds to $\til{G}_0(\til{\la})$.
However, there is some indication of this correspondence from the 
behavior of the spectral form factor.
It is known that at late times the connected part
is dominant in $\text{SFF}(x)$ and it becomes constant,
which is usually called plateau, beyond some critical value of $x$ 
\begin{equation}
\begin{aligned}
 \text{SFF}_{\text{conn}}(x)=N,\quad(x\geq x_c),
\end{aligned} 
\end{equation}
where $x_c$ is given by \cite{Brezin1997,Okuyama:2018gfr}
\begin{equation}
\begin{aligned}
 x_c=4N.
\end{aligned} 
\end{equation}
The corresponding critical time $w_H$, so-called
Heisenberg time, is given by
\begin{equation}
\begin{aligned}
 \frac{g_sw_H^2}{\pi}=x_c~~\Rightarrow~~w_H=\rt{\til{\la}}.
\end{aligned} 
\end{equation}
From \eqref{eq:0inst-K}, one can estimate the order of the
non-perturbative correction
\begin{equation}
\begin{aligned}
 \sinh^{-2}w_H\sim e^{-2w_H}=e^{-2\rt{\til{\la}}},
\end{aligned} 
\end{equation}
which is exactly the order of D-string instanton.
This suggests the relation between 
$\cG_{\text{conn}}^{(\text{0-inst})}$
and $\til{G}_0(\til{\la})$.
It would be interesting to understand this relation better.

\section{Richardson extrapolation}\label{app:richardson}
In the numerical study of the asymptotic behavior of 
a sequence, 
one can use the 
standard technique
of Richardson extrapolation
which accelerates the convergence of 
sequence towards the leading asymptotics. 
Given a sequence $\{S_m\}_{m=1,2,\ldots}$
\begin{align}
S_m=s_0+\frac{s_1}{m}+\frac{s_2}{m^2}+\cdots,\qquad
\lim_{m\to\infty}S_m=s_0,
\end{align}
its $k$-th Richardson transform is defined as
\begin{align}
S_m^{(k)}:=\sum_{n=0}^k\frac{(-1)^{k+n}(m+n)^nS_{m+n}}{n!(k-n)!}.
\label{eq:Rich}
\end{align}
After this transformation
the subleading terms in $S_m$ are canceled up to $m^{-k}$,
i.e.~$S_m^{(k)}=s_0+{\cal O}(m^{-k-1})$ and hence
the sequence $S_m^{(k)}$ has a much faster convergence to $s_0$
compared to the original $S_m$.
However, we lose some data 
in this transformation as a cost of the faster convergence:
if we know the original sequence $S_m$ up to $m=m_{\text{max}}$,
the data of $k$-th Richardson transform $S_m^{(k)}$
in \eqref{eq:Rich} 
are available only up to $m=m_{\text{max}}-k$.

One can apply the Richardson extrapolation to estimate
the large order behavior of the perturbative series
\begin{equation}
\begin{aligned}
 G\sim\sum_{m=0}^\infty a_m z^m.
\end{aligned} 
\end{equation}
When this series is Borel non-summable, there is an ambiguity
of the form
\begin{equation}
\begin{aligned}
 \lap G=\pi\ri z^{-\frac{b}{2}}e^{-\frac{A}{\rt{z}}}\sum_{n=0}^\infty f_n z^n.
\end{aligned} 
\end{equation}
Then the large $m$ behavior of $a_m$ is related to $\lap G$ as
\begin{equation}
\begin{aligned}
 a_m&\sim \int_0^\infty \frac{dz}{2\pi\ri z^{m+1}} \lap G 
=\sum_{n=0}^\infty f_n A^{-2m-b+n}\Ga(2m+b-n).
\end{aligned} 
\label{eq:am-asymp}
\end{equation}
As explained in \cite{Marino:2007te}, 
one can extract $A,b,f_n$ by constructing some sequences which converge to
them. First, $A$ is extracted from the sequence
\begin{equation}
\begin{aligned}
 A_m=2m\rt{\frac{a_m}{a_{m+1}}}.
\end{aligned} 
\end{equation}
From \eqref{eq:am-asymp} one can see that
$A_m$ approaches $A$ as $m\to\infty$. In practice, $A$ is estimated numerically
by applying the Richardson extrapolation to the sequence $\{A_m\}_{m=1,2,\ldots}$.
Once we find $A$, then we can extract $b$ from the sequence
\begin{equation}
\begin{aligned}
 b_m=m\left(A^2\frac{a_{m+1}}{4m^2a_m}-1\right)-\hf,
\end{aligned} 
\end{equation}
which converges to $b$ in the limit $m\to\infty$.
Finally, $f_n$ can be found from the sequence
\begin{equation}
\begin{aligned}
 f_{n,m}=\frac{1}{A^{-2m-b+n}\Ga(2m+b-n)}\left[a_m-\sum_{k=0}^{n-1}
f_kA^{-2m-b+k}\Ga(2m+b-k)\right],
\end{aligned} 
\label{eq:fnm}
\end{equation}
which converges to $f_n$ as $m\to\infty$. 
For instance, $f_0$ can be obtained from the large $m$ limit of $f_{0,m}$ given by
\begin{equation}
\begin{aligned}
 f_{0,m}=\frac{a_m}{A^{-2m-b}\Ga(2m+b)}.
\end{aligned} 
\end{equation}
Once we know $f_k~(0\leq k\leq n-1)$, 
one can extract $f_n$ numerically 
from the large $m$ asymptotics of the sequence $\{f_{n,m}\}_{m=1,2,\ldots}$
using the Richardson extrapolation.

\bibliography{refs}
\bibliographystyle{utphys}

\end{document}